\documentclass[10pt, a4paper]{deepmind}

\usepackage{times}

\usepackage{xspace}
\usepackage{siunitx}
\usepackage{amsmath}
\usepackage{amssymb}
\usepackage{enumitem}
\usepackage[T1]{fontenc}
\usepackage{multicol}
\usepackage{caption}
\usepackage{subcaption}
\usepackage{sidecap}
\usepackage{multirow}
\usepackage{placeins}
\usepackage{titlesec}
\usepackage[capitalise,noabbrev]{cleveref}
\usepackage{graphicx}
\usepackage{url}
\usepackage{lineno}
\usepackage{titletoc}
\usepackage{bibunits}
\usepackage{bm}
\usepackage{array}
\usepackage{placeins}

\makeatletter
\usepackage{etoolbox}
\patchcmd\H@refstepcounter{\protected@edef}{\protected@xdef}{}{}
\makeatother

\ifdefined\unit\else
  \ifdefined\NewCommandCopy
    \NewCommandCopy\unit\si
  \else
    \NewDocumentCommand\unit{O{}m}{\si[#1]{#2}}
  \fi
\fi

\crefformat{footnote}{#2\footnotemark[#1]#3}

\crefname{appsec}{Appendix Section}{Appendix Sections}
\crefname{appfig}{Appendix Figure}{Appendix Figures}
\crefname{apptab}{Appendix Table}{Appendix Tables}
\crefname{appeq}{Appendix Equation}{Appendix Equations}

\makeatletter
\def\varlevel#1{\def\tempa{\lowercase{#1}}\futurelet\next\varlevel@i}%
\def\varlevel@i{\ifx\next\bgroup\expandafter\varlevel@ii\else\expandafter\varlevel@end\fi}%
\def\varlevel@ii#1{\textsc{\tempa#1}}%
\def\varlevel@end{\textsc{\tempa}}%
\makeatother

\makeatletter
\def\enotation#1{\def\tempa{#1}\futurelet\next\enotation@i}%
\def\enotation@i{\ifx\next\bgroup\expandafter\enotation@ii\else\expandafter\enotation@end\fi}%
\def\enotation@ii#1{$\tempa\mathrm{e}{#1}$}%
\def\enotation@end{$\tempa$}%
\makeatother

\title{MAUSAM: An Observations-focused assessment of Global AI Weather Prediction Models During the South Asian Monsoon}

\author[1,*]{Aman Gupta}
\author[1]{Aditi Sheshadri}
\author[2]{Dhruv Suri}

\affil[1]{Department of Earth System Science, Stanford University, United States}
\affil[2]{Department of Energy Science \& Engineering, Stanford University, United States}

\correspondingauthor{ag4680@stanford.edu}

\begin{abstract}

Accurate weather forecasts are critical for societal planning and disaster preparedness. Yet these forecasts remain challenging to produce and evaluate, especially in regions with sparse observational coverage. Current evaluation of artificial intelligence (AI) weather prediction relies primarily on reanalyses, which can obscure important deficiencies due to prevailing biases in reanalyses themselves. Here we present MAUSAM (Measuring AI Uncertainty during South Asian Monsoon), an evaluation of seven leading AI-based forecasting systems - FourCastNet, FourCastNet-SFNO, Pangu-Weather, GraphCast, Aurora, AIFS, and GenCast - during the South Asian Monsoon, using ground-based weather stations, rain gauge networks, and geostationary satellite imagery. The AI models demonstrate impressive forecast skill during monsoon across a broad range of variables, ranging from large-scale surface temperature and winds to precipitation and cloud cover. The models also generate realistic zonal means and eddy statistics on subseasonal to seasonal timescales, highlighting the strength of purely data-driven weather prediction. However, the models still exhibit systematic errors in finer-scale features like the underprediction of extreme precipitation ("tail" events), divergent cyclone tracks, and the mesoscale kinetic energy spectra, highlighting avenues for future improvement. A comparison against a combination of in situ and remote sensing observations reveals forecast errors up to 15–45\% larger than those relative to reanalysis and traditional forecasts, indicating that reanalysis-centric benchmarks can overstate forecast skill. Of the models assessed, ECMWF's AIFS achieves the most consistent representation of atmospheric variables, with GraphCast and GenCast also showing strong skill. The analysis presents a framework for systematically evaluating global AI weather models on regional prediction, and the results highlight both the promise and current limitations of AI weather prediction in data-sparse regions. This underscores the importance of observation-based evaluation for future operational adoption.

\end{abstract}

\begin{document}
\maketitle

\section*{Introduction}

Recent progress in global weather prediction has been dominated by the emergence of artificial intelligence (AI)-driven weather prediction models. Advances in AI have inspired the development of fast and inexpensive global weather emulators that reduce a complex numerical task to a data-driven task that can be accomplished within minutes on a single GPU, i.e., up to a factor of 10,000-50,000 faster. Pioneering efforts in this rising field include \cite{Scher2018, Dueben.Bauer2018}, DLWP \cite{Weyn.etal2021}, FourCastNet \cite{Pathak.etal2022}, Pangu-Weather \cite{Bi.etal2023}, and GraphCast \cite{Lam.etal2023}. More recent models including, but not limited to, GenCast \cite{Price.etal2025}, NeuralGCM \cite{Kochkov.etal2024}, CorrDiff \cite{Mardani.etal2024}, and AIFS \cite{Lang.etal2024}, have been developed to generate both deterministic and probabilistic (ensemble) forecasts. 

Concurrently, growing cognizance of the limited applicability and reanalysis-focused training of these models has also led to the creation of multiple foundation models, which aim to be used (in addition to weather forecasting) for multiple ``downstream" applications, including air quality prediction, burn scar detection, hurricane track and intensity prediction, wave prediction, etc. Some notable foundational models include Prithvi HLS \cite{Jakubik.etal2023}, Prithvi WxC \cite{Schmude.etal2024}, Aurora \cite{Bodnar.etal2025}, and AtmoRep \cite{Lessig.etal2023}.

Practically all AI weather prediction models (hereafter AIWP models)
are, either fully or in part, trained on the global reanalysis dataset, ERA5 \cite{Hersbach.etal2020}, produced by the European Centre for Medium-Range Weather Forecasts (ECMWF), owing to its high spatio-temporal resolution and multi-decadal coverage. Some models, including GraphCast and PanguWeather, have both \emph{default} versions, trained on ERA5, and \emph{operational} versions, which are subsequently fine-tuned on forecasts from ECMWF's 9 km High Resolution Forecast Model (HRES). As an upside, this allows AIWP models to bypass numerical integrations of the equations of fluid flow, and instead, directly learn weather evolution from global weather data. However, as a downside, heavy reliance on a single dataset puts an upper ceiling on the trained AI model's skill and resolution. Despite incorporating over $10^7$ observation points daily through sophisticated data assimilation algorithms, ERA5 relies on time-marching the earth system initialization state using a forecasting model. As a result, ERA5 itself could be influenced by data assimilation errors, parameterization drifts, and numerical biases present in the underlying model, i.e., ECMWF's Integrated Forecasting System (IFS) \cite{AdrianSimmons2020,Truong.etal2022,  Wilczak.etal2024, Wolf.etal2025}.


Multiple recent deterministic AIWP models have outperformed conventional numerical weather prediction (NWP) models on key error metrics, including global root mean square error (RMSE), mean absolute error (MAE), and Anomaly Correlation Coefficient (ACC) (for instance, \cite{Lang.etal2024}). Likewise, probabilistic models like GenCast demonstrate a better continuous ranked probability score (CRPS) than ECMWF's IFS ensemble over all lead times \cite{Price.etal2025}. 

Beyond testing on RMSE-like metrics, multiple studies have tested AIWP models' performance around key extreme weather events, revealing a systematic underestimation of wind speeds during cyclones \cite{Charlton-Perez.etal2024}, and reduced predictability around Category-5 cyclones \cite{Sun.etal2025, Bouallègue.etal2024} and extreme precipitation \cite{Sun.etal2025b}. At the same time, conflicting conclusions from independent studies have also highlighted both their high skill in representing the large-scale physics \cite{Hakim.Masanam2024, Hua.etal2025}, and poor skill in representing the small-scale dynamics and surface physics \cite{Bonavita2024, Olivetti.Messori2024}, making them potentially suitable for at least some scale-selective dynamical analysis. 

\begin{figure}
   \centering
   \includegraphics[width=\linewidth]{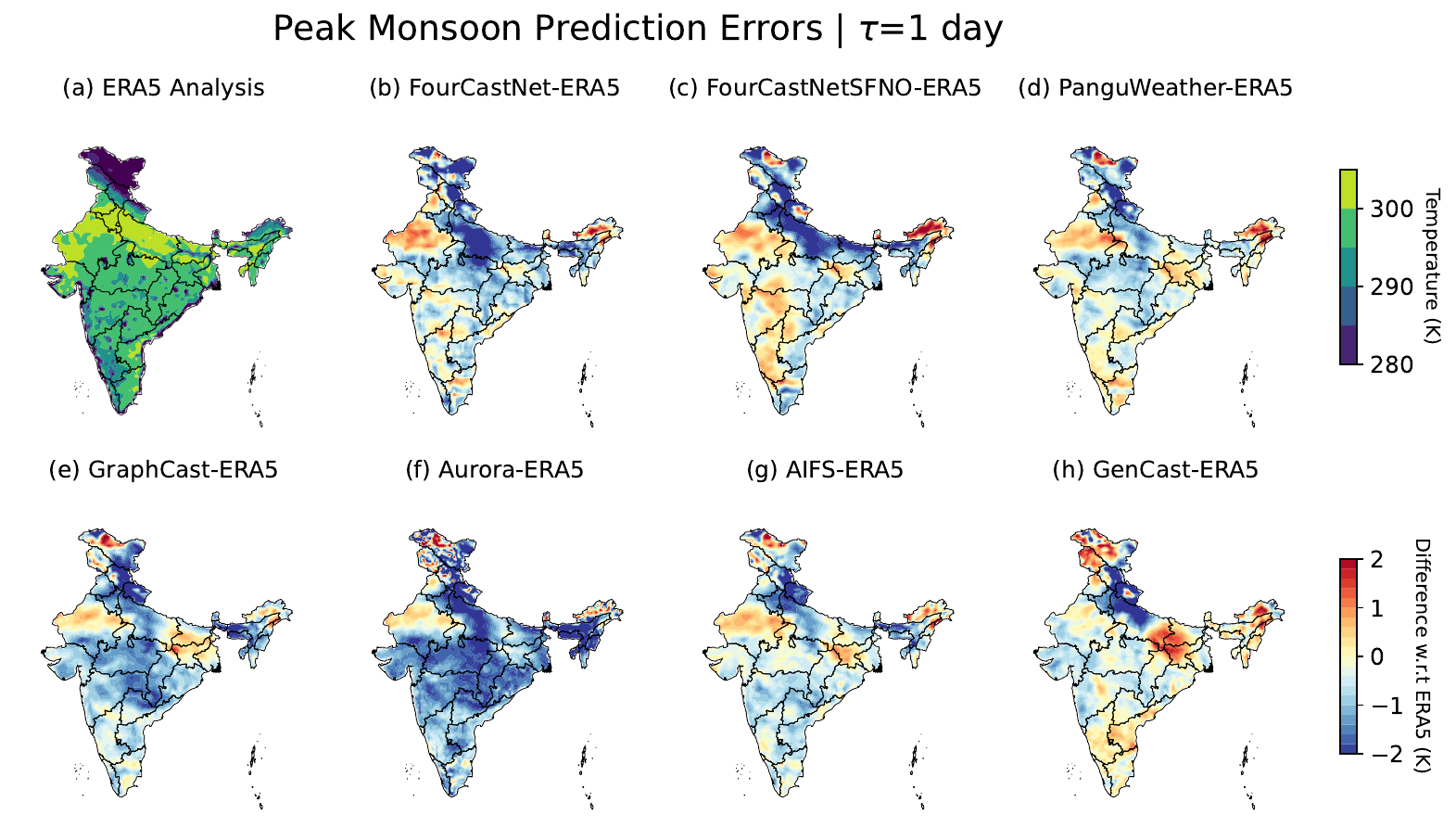}
   \caption{\small \textbf{Systematic biases in AI weather models during peak monsoon conditions.} Spatial distribution of 2-meter temperature for ERA5 analysis (a) and prediction errors relative to ERA5 for seven AI weather models (b-h) at $\tau = 1$ day lead time, initialized on 15 July 2022 during peak monsoon activity. Contour intervals are 3 K for ERA5 temperatures and 1 K for model-analysis differences. All AI models exhibit characteristic cold biases over the Indo-Gangetic Plain and warm biases over the Western Ghats, with error magnitudes of 1-2 K that reveal systematic deficiencies in monsoon thermodynamic representation. GraphCast (e), AIFS (g), and GenCast (h) show the smallest systematic biases, while FourCastNet variants (b,c) and Aurora (f) exhibit larger regional errors.} 
   \label{fig:monsoon_errors}
\end{figure}

The inter-model variability among AIWP models becomes apparent when examining forecasts during dynamically active periods like the Monsoon. Figure \ref{fig:monsoon_errors} illustrates substantial differences in day-ahead surface temperature predictions across seven state-of-the-art AIWP models during peak monsoon conditions over the Indian subcontinent on 15 July 2022. Forecast differences of several degrees Celsius are noted among the models, despite identical initialization using ERA5. The global differences are shown in Figure \ref{fig:global_errors}. These disparities highlight the sensitivity of AI-based forecasting systems to architectural choices and training methodologies, while simultaneously revealing the inadequacy of reanalysis-based validation.

Findings such as these inspire the need for continuing evaluation and benchmarking \cite{Rasp.etal2024, Brenowitz.etal2025a} of AIWP models over varied geographies and weather conditions. Most existing benchmarking initiatives test AIWP models against modeling output, leading to a circular paradigm where AIWP models are evaluated against the traditional model data they are trained on, obscuring their performance against actual ground-based and remote sensing observations that often drive operational decisions.

This study focuses on evaluating the forecast of critical weather variables from seven advanced AIWP models against ground-based and remote-sensing observations (458 weather stations + rain gauge data + geostationary satellites). The models (as described in the Methods section) include FourCastNet, FourCastNet-SFNO, Pangu-Weather, GraphCast, Aurora, AIFS, and GenCast (both deterministic and ensemble versions). We compare both the global power spectra and regional forecasting errors over the Indian subcontinent during the South Asian Monsoon. This allows (a) connecting global model errors with regional model errors, and (b) testing model predictability in the tropics (where the dynamical evolution of the atmosphere is likely governed by an interplay of fast dynamics, moist convection, land-ocean coupling, and topographical effects) during one of the most dynamic periods of the year. The models are tested against observations of surface temperature, wind speeds, extreme precipitation, and cloud cover, to obtain an observationally validated sense of the models' performance.

\section*{Results}

First, we conduct a global error analysis of the different models for the year 2022 over short-to-medium-range lead times. Then we sequentially compare the predictions of the kinetic energy spectrum, temperature power spectrum, precipitation predictions, cloud cover, and cyclone trajectories in AIWP models against a combination of the most relevant observations, traditional weather models, and/or reanalysis. In doing so, we first focus on global errors and energy balance and then refine our focus towards the Indian subcontinent.

\subsection*{Moving from reanalysis-focused to observation-focused evaluation of AIWP models}

To get a more precise sense of their surface prediction skill, we compared the AIWP models against both ERA5 and observations from ground-based weather stations. We compared mean absolute errors (MAE) of surface temperature, zonal wind, and meridional wind predictions over the Indian subcontinent for the year 2022 and present them in a style similar to WeatherBench \cite{Rasp.etal2024} (their Figs. 1-3). The MAE is computed against observations from 458 Indian Meteorological Department ground-based weather stations, ERA5 reanalysis, HRES predictions, and IFS ensemble mean predictions.

\begin{figure}
   \centering
   \includegraphics[width=\linewidth]{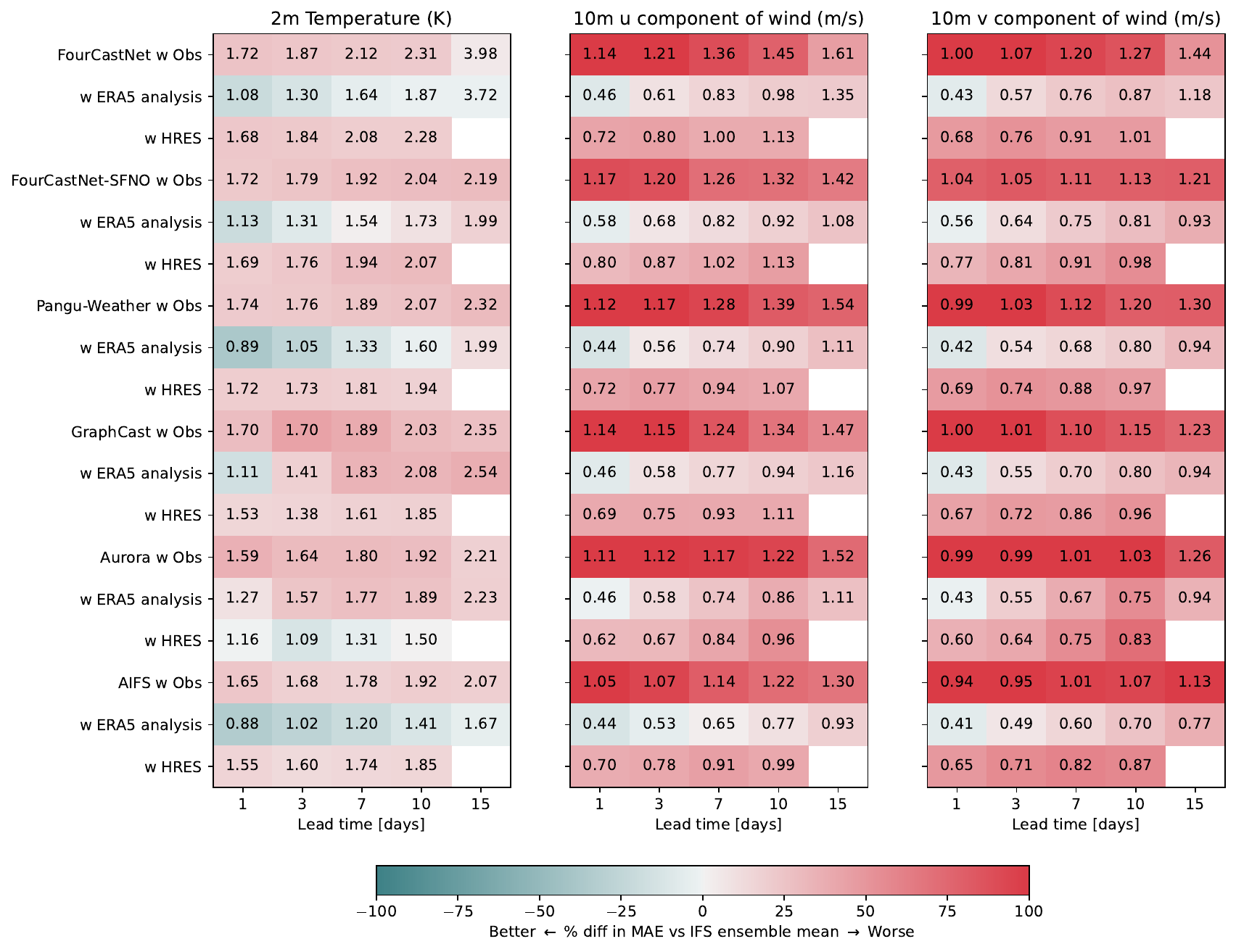}
   
   \caption{\small \textbf{Observational validation exposes systematic biases masked by reanalysis-centric evaluation.} Comparison of mean absolute errors for AI weather models evaluated against Indian Meteorological Department observations, ERA5 analysis, and ECMWF operational forecasts (HRES) for 2-meter temperature (left), 10-meter u-component wind (center), and 10-meter v-component wind (right) during the year 2022. Error values are shown for various forecast lead times. The color scale indicates the percentage difference in mean absolute error relative to ECMWF's IFS ensemble mean performance, with negative values (green) indicating better performance and positive values (red) indicating worse performance. All AI models show substantially larger errors when validated against observations compared to ERA5-based evaluation, with discrepancies increasing systematically with lead time.}
   \label{fig:weatherbench_comparison}
\end{figure}

For all models except Aurora, the MAE is found to be the lowest w.r.t. ERA5 (Figure \ref{fig:weatherbench_comparison}) for all variables. This is no surprise, given that all these models were either exclusively trained or pre-trained on ERA5. Aurora, an exception, was trained on sixteen different model outputs, ranging from ERA5, HRES, and IFS ensemble, to NOAA's GFS and CMIP6 output, and NASA's MERRA2 reanalysis, and produces the lowest MAE w.r.t. HRES for all variables. For up to 1-day ahead forecasts, GraphCast has the lowest errors against ERA5, but at longer lead times, the errors are lowest against HRES, most likely due to fine-tuning on five years of HRES predictions.

The models consistently have the highest errors against the 458 weather station observations, with MAE exceeding those w.r.t. ERA5 by 15-45\% across all meteorological variables. For skin temperature, the predictions from AIFS and Aurora are found to be the most accurate (Figure \ref{fig:weatherbench_comparison}, left column). For surface zonal wind, FourCastNet and GraphCast offer the least prediction error over all lead times (Figure \ref{fig:weatherbench_comparison}, middle column). Finally, for the surface meridional wind, AIFS provides the most accurate predictions over all lead times (Figure \ref{fig:weatherbench_comparison}, right column). For a two-week ahead forecast ($\tau=15$ days), the errors against observations differ from the errors against ERA5 by a factor of approximately 2. Similar patterns in MAE are observed for temperature at both 850 hPa (Figure \ref{fig:weatherbench850hpa}) and 500 hPa (Figure \ref{fig:weatherbench500hpa}). 

\subsection*{Intercomparison of Temperature, Kinetic Energy, and Moisture Spectrum Across AIWP Models}

To better understand the scales at which these errors exist, we analyzed the eddy kinetic energy (EKE) spectrum, the temperature power spectrum, and the specific humidity spectrum at the surface and in the free atmosphere. 

\begin{figure}
   \centering
   \includegraphics[width=0.8\linewidth]{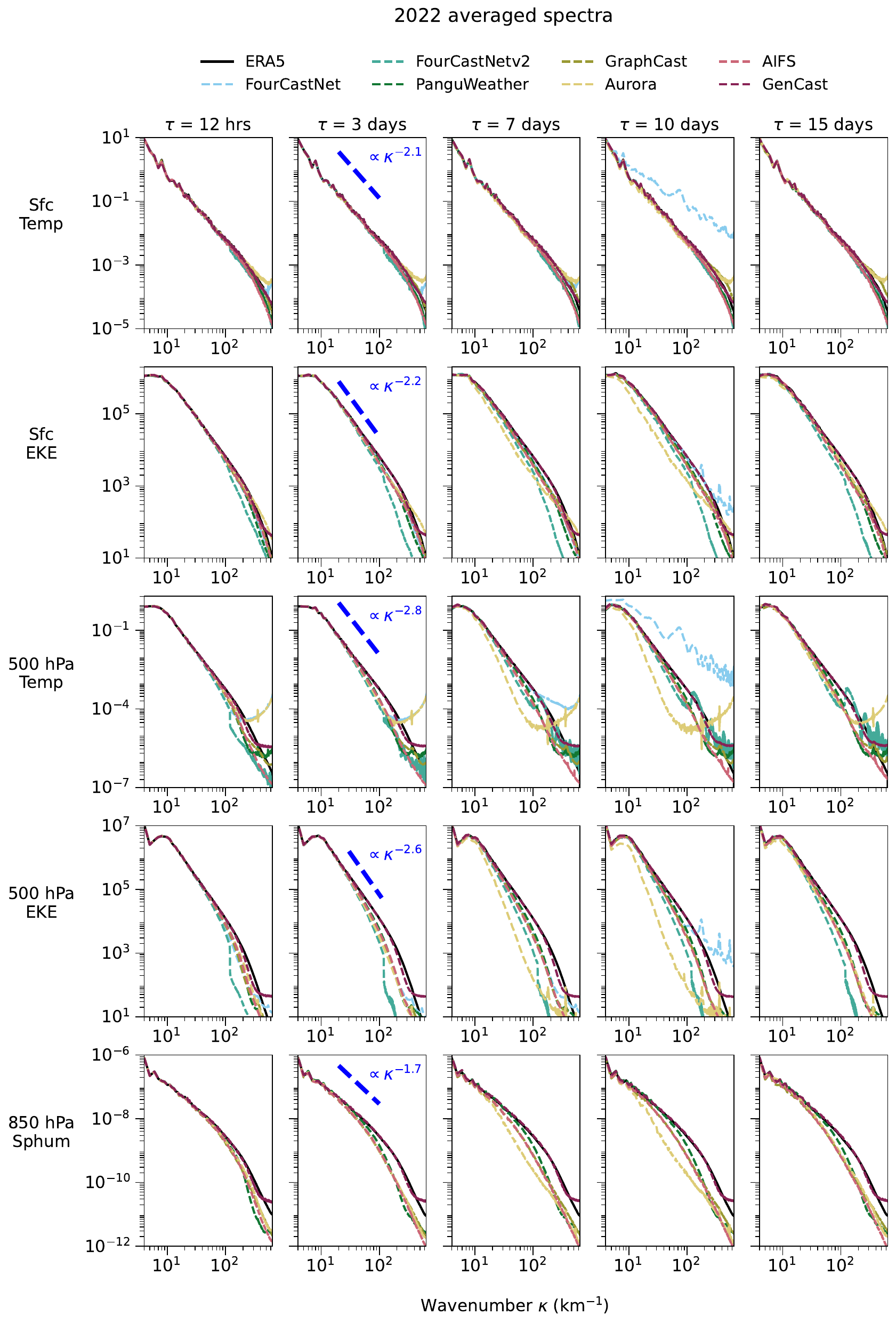}
   \caption{\small \textbf{Spectral biases across atmospheric levels reveal limitations in synoptic- and meso-scale representation.} Power spectral density comparison for multiple atmospheric variables during the 2022 monsoon season, showing surface temperature (top row), surface eddy kinetic energy (second row), 500-hPa temperature (third row), 500-hPa eddy kinetic energy (fourth row), and 850-hPa specific humidity (bottom row) across forecast lead times from $\tau = 12$ hours to $\tau = 15$ days. Wavenumber $\kappa$ is expressed in km$^{-1}$. All AI models (colored lines) are compared against ERA5 analysis (black line). Systematic energy deficits emerge at mesoscales ($k \approx 10^1$--$10^2$ km$^{-1}$) with increasing severity at longer lead times.}
   \label{fig:spectrum_2022}
\end{figure}

The different models agree strongly, amongst themselves and with ERA5, on the near-surface temperature spectrum across planetary-, synoptic-, and meso-scales (Figure \ref{fig:spectrum_2022}; first row). GenCast closely matches the ERA5 spectrum at all scales and lead times. Aurora exhibits more power at finer scales, while all the other models exhibit weaker power for wavenumbers 100 and higher (approximately corresponding to scales 300-400 km and shorter). The most notable exception is the FourCastNet model, which tends to ``blow-up" and generate unphysically large values after 9-10 days of autoregressive rollout. The ``blow-up" starts over the poles after 9 days (not shown) and then spreads globally.

Most models produce a physically consistent spectral slope between $\propto\kappa^{-2}$ to $\propto\kappa^{-3}$ for the EKE spectrum (Figure \ref{fig:spectrum_2022}; second row), indicating a reasonably skillful emulation of energy transfers across spatial scales. However, similar to the surface temperature power spectrum, while the EKE spectrum for GenCast nearly identically matches the EKE spectrum from ERA5 across scales, almost all the other models underestimate the power over a range of wavenumbers, even for short lead times. For short-range lead times ($\tau$ = 12 hrs to 3 days), the underestimation is most severe for FourCastNet-SFNO. However, for longer lead times ($\tau$ = 7 and 10 days), forecasts from Aurora exhibit a notable weak bias which pervades even the planetary scales. These spectral anomalies are, however, corrected for by $\tau$=15 days. This behavior is quite likely due to Aurora being trained on multiple datasets over different lead times. All models, except FourCastNet, FourCastNet-SFNO, and Aurora, project consistent behaviour across all lead times.

Similar but amplified patterns are found higher up in the atmosphere. At 500 and 850 hPa (Figure \ref{fig:spectrum_2022}; third and fourth row), the power spectrum from GenCast agrees well with ERA5's power spectrum, but the power accumulation at near-grid scales is higher than it is at the surface. Pangu-Weather, GraphCast, and AIFS have even weaker power (than ERA5) across all wavenumbers than at the surface. Similarly, the spectra from FourCastNet, FourCastNet-SFNO, and Aurora continue to project spurious perturbations at wavenumbers 100 and higher. Aurora continues to critically underestimate the spectral power over the synoptic scales but overestimates it over finer scales for $\tau$=7 and 10 days. Finally, all models except GenCast underestimate near-surface specific humidity to a similar degree for all lead times (Figure \ref{fig:spectrum_2022}; fifth row).

More simply put, GenCast has the least error in spectral representation w.r.t. ERA5 both in the surface and in the free atmosphere. Next, GraphCast, PanguWeather, and AIFS offer very similar and, arguably, physically consistent performance across scales. FourCastNet provides a comparative performance for short lead times before growing unstable around 9-10 days. FourCastNet-SFNO tends to underestimate power at small scales, and this underestimation is higher for higher lead times. Finally, Aurora performs well for short lead times, but begins to demonstrate an unphysical distribution of energy across scales after 7 days. 

Similar patterns are found for the regional power spectra over the Indian Subcontinent during the peak Monsoon season, June - September 2021-2024 (Figure \ref{fig:spectrum_multiyear}). The underestimation is much less severe, indicating that the regional spectral deviations across models are potentially more severe elsewhere.

\subsection*{Precipitation validation with Monsoon Rain Gauge Observations}

Does an underestimation of the specific humidity power at small scales by AIWP models equate to an underestimation of local precipitation by them? To investigate this, we analyzed the daily-averaged day-ahead precipitation predictions from GraphCast, AIFS, and GenCast, and validated them against ERA5 and ground-based rain gauge station data from IMD \cite{Pai.etal2014} from the onset of the Monsoon (end of May) to its retreat (beginning of October).

\begin{figure}
   \centering
   \includegraphics[width=\linewidth]{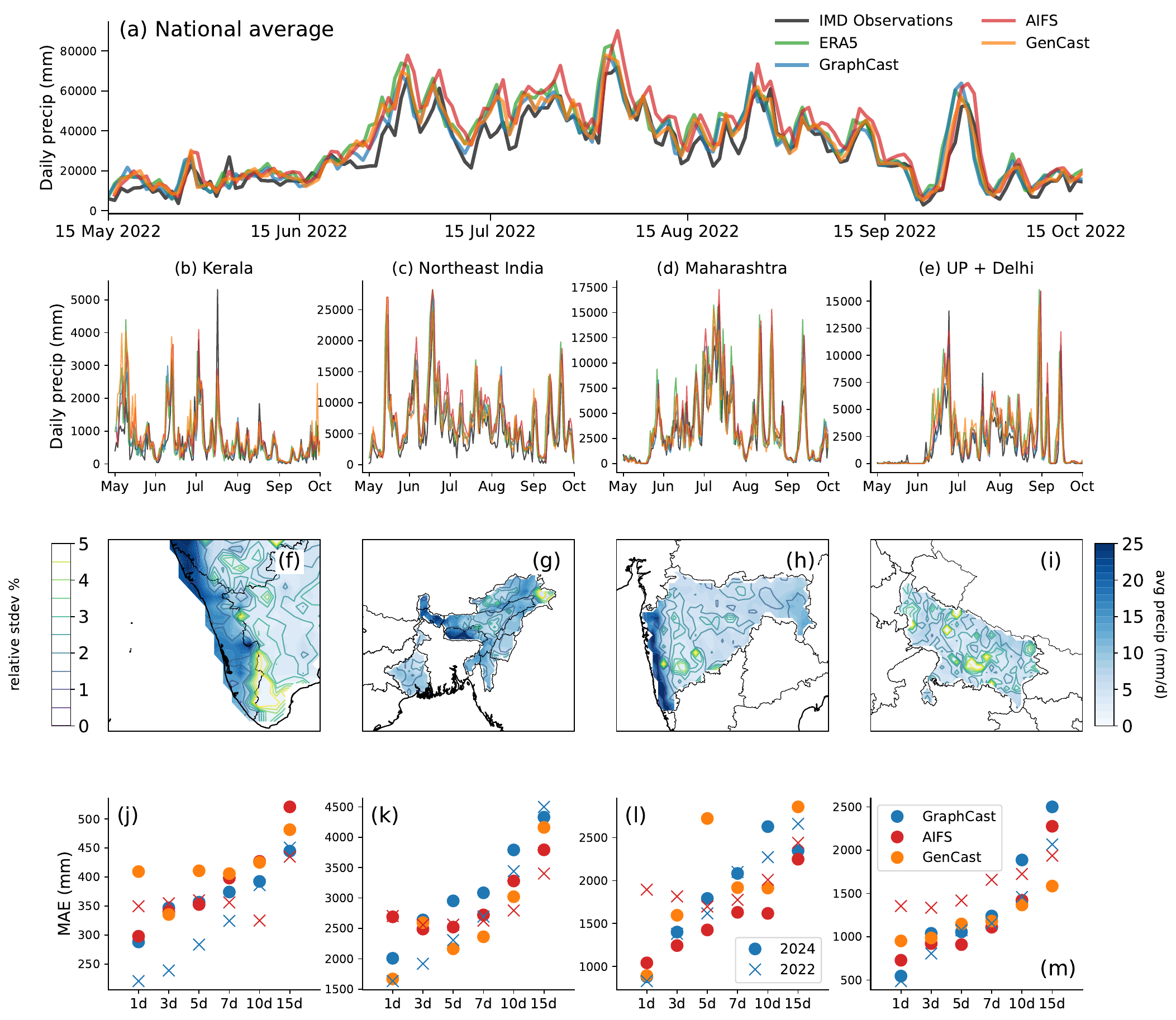}
   \caption{\small \textbf{Regional precipitation time series reveal underestimation of extreme precipitation.} Daily precipitation accumulations during the 2022 monsoon season for (a) national average, (b) Kerala, (c) Northeast India, (d) Maharashtra, and (e) Uttar Pradesh plus Delhi, comparing IMD observations (black), ERA5 reanalysis (gray), and AI weather models (colored lines). Lower panels show mean absolute error as a function of forecast lead time for each region, with separate lines for 2024 and 2022 seasons where available. (f)-(i) Individual model standard deviations were computed w.r.t. IMD. (j)-(m) are computed w.r.t. IMD. Crosses and dots denote errors for the 2022 and 2024 Monsoon periods, respectively.}
   \label{fig:precip_timeseries}
\end{figure}

India receives as much as 80\% of its annual rainfall during the summer Monsoon. The integrated daily precipitation (from day-ahead forecasts) from the three models agrees strongly with precipitation from ERA5 and that recorded by IMD rain gauges over the whole country for the whole season. All models exhibit a slight wet-bias throughout the season (Figure \ref{fig:precip_timeseries}(a)). The spatial structure of these prediction differences w.r.t. both ERA5 and IMD gridded precipitation data for years 2024 and 2022 is shown in Figure \ref{fig:precip_2024} and Figure \ref{fig:precip_2022}, respectively. The wet bias could potentially be due to an underestimation of rainfall by rain gauges due to varying gauge types, oversplash, strong winds, and limited coverage. Most of these factors should not affect ERA5 because it assimilates rainfall from multiple microwave soundings with better spatial coverage. 

The wet-bias can also be noted in the state-wise precipitation time series for the states of (in the order of monsoon onset) Kerala, Northeastern states, Maharashtra, and Uttar Pradesh+Delhi (Figure \ref{fig:precip_timeseries}(b)-(e)). GenCast reports the highest magnitude of rainfall over Kerala (Southern India). Here, all models substantially overestimate the precipitation in mid-May (during the onset) and the end of June. However, the models collectively fail to predict the extreme precipitation towards the end of July as indicated by IMD data (Figure \ref{fig:precip_timeseries}(b)). Northeast India, comprising the state of Meghalaya, is one of the wettest regions in the world. Here, the models exhibit some wet bias but overall predict the day-ahead intermittent rainfall patterns remarkably well throughout the monsoon period (Figure \ref{fig:precip_timeseries}(b)). The wet bias is the least for Maharashtra (Western Indian).

The regions with the highest intermodel differences in precipitation predictions are shown in Figures \ref{fig:precip_timeseries}(f)-(i). For each state, the maximum relative deviation among models is found away from the wettest regions within the state. In Kerala, the models' standard deviation is the highest around its eastern boundary (Figure \ref{fig:precip_timeseries}(f)). In the northeast, the models diverge the most over eastern Arunachal Pradesh (Figure \ref{fig:precip_timeseries}(g)). Over Maharashtra, the models have the highest errors in the southern parts of the state, away from the western coast (Figure \ref{fig:precip_timeseries}(h)). Uttar Pradesh + Delhi doesn't have a clearly defined wettest region, and the deviations are spread more uniformly (Figure \ref{fig:precip_timeseries}(i)).

All AIWP models provide solid predictions of the three-day ahead nationally-averaged rainfall (Figure \ref{fig:precip_timeseries_lag3}(a)), but notable errors emerge at regional levels. GenCast's performance degrades the quickest. Most notably, for 3-day ahead precipitation forecasts, GenCast misses most precipitation peaks over Kerala, and overpredicts peak precipitation over Maharashtra (Figure \ref{fig:precip_timeseries_lag3}(b)-(e)). 

During 2024 Monsoon, GraphCast provides the most accurate day-ahead rainfall predictions over all four states (blue dots in Figures \ref{fig:precip_timeseries}(j)-(m)), but AIFS (red dots) outperforms GraphCast over longer lead times. For 2022 (crosses), however, GraphCast consistently offers lower errors for most lead times than AIFS. Averaging over a longer period might be necessary to decisively conclude which among AIFS and GraphCast offers better precipitation predictions. Both models employ a graph neural network-based architecture \cite{Scarselli.etal2009}, which might explain similar performance. 

\begin{figure}
   \centering
   \includegraphics[width=0.8\linewidth]{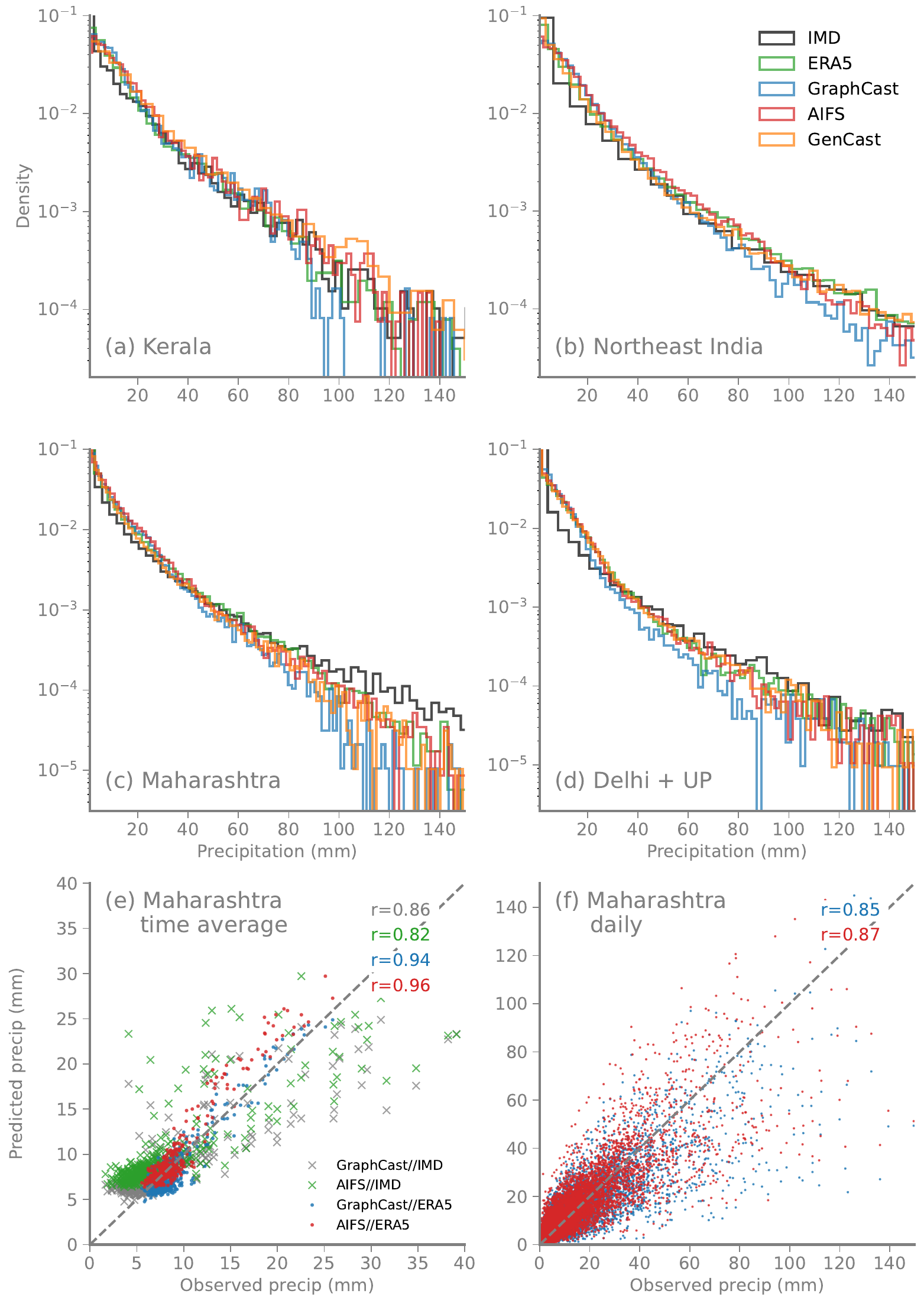}
   \caption{\small \textbf{Probability density functions expose systematic biases in representing precipitation extremes across regions.} Precipitation probability distributions for four climatologically distinct regions during the 2022 monsoon season, comparing IMD observations, ERA5 reanalysis, and AI weather models. All models systematically underrepresent the frequency of heavy precipitation events (daily accumulations $> 50$ mm) while overestimating light-to-moderate precipitation occurrence. Scatter plots for Maharashtra show (e) time-averaged seasonal relationships and (f) daily-resolution comparisons between predicted and observed precipitation. The systematic underestimation of extreme precipitation events has critical implications for flood forecasting and water resource management applications where accurate representation of the upper tail of the precipitation distribution is essential.}
   \label{fig:precip_distributions}
\end{figure}

The most likely cause of regional wet-bias could be an underestimation of rainfall by the IMD rain gauges. Further analysis of such underestimation is beyond the scope of this study, but underestimation due to lower-than-ideal rain gauge density and improper interpolation would be consistent with conclusions drawn by past studies \cite{Jena.etal2020}. 

Precipitation distributions from the three models agree closely with the precipitation distribution from ERA5. However, comparison with 2024 raingauge observations reveals model tendencies to overestimate weak rainfall events and underestimate moderate-to-strong rainfall events (Figure \ref{fig:precip_distributions}(a)-(d)). Despite providing the most accurate precipitation estimates over different lead times, GraphCast notably underestimates extreme precipitation (tail) in all four states. AIFS's tails agree the strongest with both reanalysis and IMD observations. All models poorly emulate the precipitation tails in Maharashtra, with AIFS being the closest to both ERA5 and IMD (Figure \ref{fig:precip_distributions}(d)).

Nationwide and state-wise precipitation analysis reveals a systematic underestimation of precipitation by GraphCast (relative to AIFS) w.r.t. both IMD observations and ERA5. This can be seen clearly in Figure \ref{fig:precip_distributions}(e) where the dots and crosses represent a comparison of time-averaged precipitation with ERA5 and IMD, respectively. Since both models were trained on ERA5, their predictions are more strongly correlated with ERA5 than with IMD, even as ERA5 does not assimilate any precipitation data from India. For daily predictions, the underestimation is much weaker than is suggested by the time means (dots in Figure \ref{fig:precip_distributions}(e) vs (f)).

As will be discussed in the following subsection, since AIWP models generate 6-hourly forecasts only, they tend to miss a major fraction of high-frequency precipitation variability, highlighting a major limitation of these models in representing extreme precipitation.

\subsection*{Validating Cloud Cover in AIFS against INSAT-3DS Satellite Imagery}

Cloud nowcasting and seasonal cloud prediction continue to be a challenge for weather and climate models. For weather models in particular, the lack of global high-frequency vertical cloud profiles, an inaccurate representation of cloud microphysics, and the coupling of cloud formation with other poorly constrained physical processes like convection and boundary layer turbulence impede cloud nowcasting. Motivated by the value of cloud cover predictions to solar farm operations, we assess the predicted cloud cover in AIWP models by validating them against geostationary satellite products during the Monsoon.

\begin{figure}
    \centering
    \includegraphics[width=\linewidth]{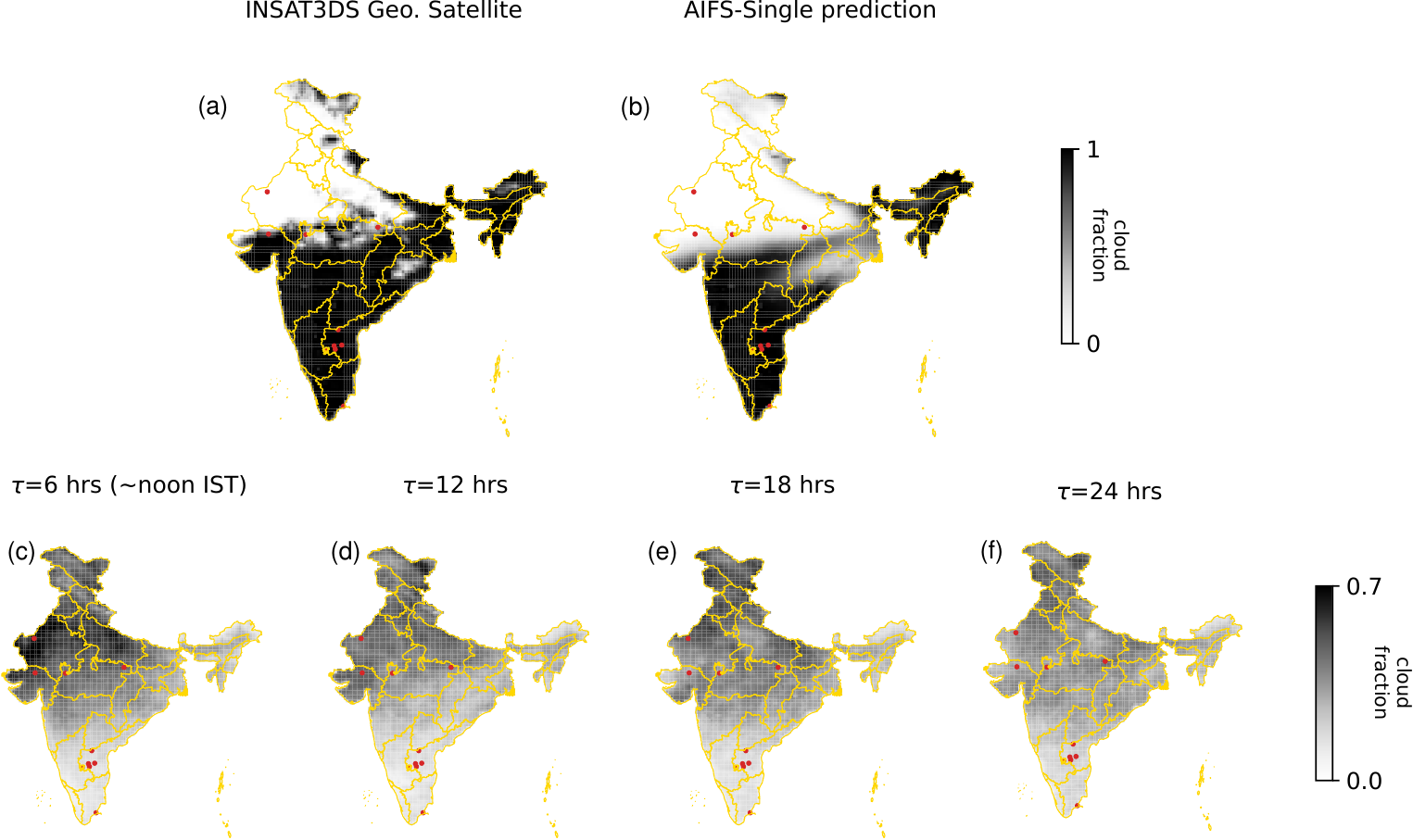}
    \caption{\textbf{Strong agreements between AIFS cloud cover and INSAT-3DS Satellite Imagery.} (a) The observed cloud cover by INSAT-3DS on 20 May 2025 1200UTC, and (b) corresponding 12-hour ahead prediction by AIFS-single AIWP model. INSAT3DS began operation on 13 July 2024, but only provides cloud fraction (Level-2 post-processed output) starting 30 September 2024. (c)-(f) Cloud cover errors for different lead times averaged between 20 September to 15 October 2024, and 15 May 2025 and 15 June 2025. Red dots indicate locations of the 10 largest solar plants in India.}
    \label{fig:clouds_insat}
\end{figure}

At present, ECMWF's AIFS is the only AIWP model that outputs cloud cover. AIFS outputs low, medium, and high cloud cover, corresponding to different atmospheric altitudes, with high mostly corresponding to icy cirrus clouds, medium mostly corresponding to altostratus and altocumulus clouds, and low corresponding to a combination of stratus, cumulus, and cumulonimbus clouds. The INSAT-3DS geostationary satellite does not directly measure the cloud cover, but infers it using a combination of cloud top pressure, cloud optical thickness, and brightness temperature. We use the processed product from INSAT-3DS to validate the total cloud cover in AIFS. The cloud cover in AIFS is converted to the total cloud cover using the random overlap method described in the Methods section.

For a selected timestamp on 20 May 2025 1200UTC, instantaneous cloud cover derived from both INSAT and AIFS shows strong agreement (Figure \ref{fig:clouds_insat}, top row). Similar features, including a full cloud cover over Southern India and Northeast India, are noted in both AIFS and INSAT-3DS with some differences in magnitude. 
The cloud cover predicted by AIFS misses certain fine-scale features over Central India, which are quite evident in INSAT observations.
Averaging over multiple days and for multiple lead times reveals maximum absolute errors between INSAT imaging and AIFS predictions during noon-time, i.e., $\tau$=6 hrs (Figure \ref{fig:clouds_insat}, bottom row). For this time of the day, the lowest errors are obtained over Southern and Northeastern India, and the highest errors for all lead times are obtained over North India (especially the states of Rajasthan, Gujarat, and Uttar Pradesh). Some of the biggest solar plants in India are located in Rajasthan and Gujarat.

\begin{figure}
   \centering
   \includegraphics[width=0.95\linewidth]{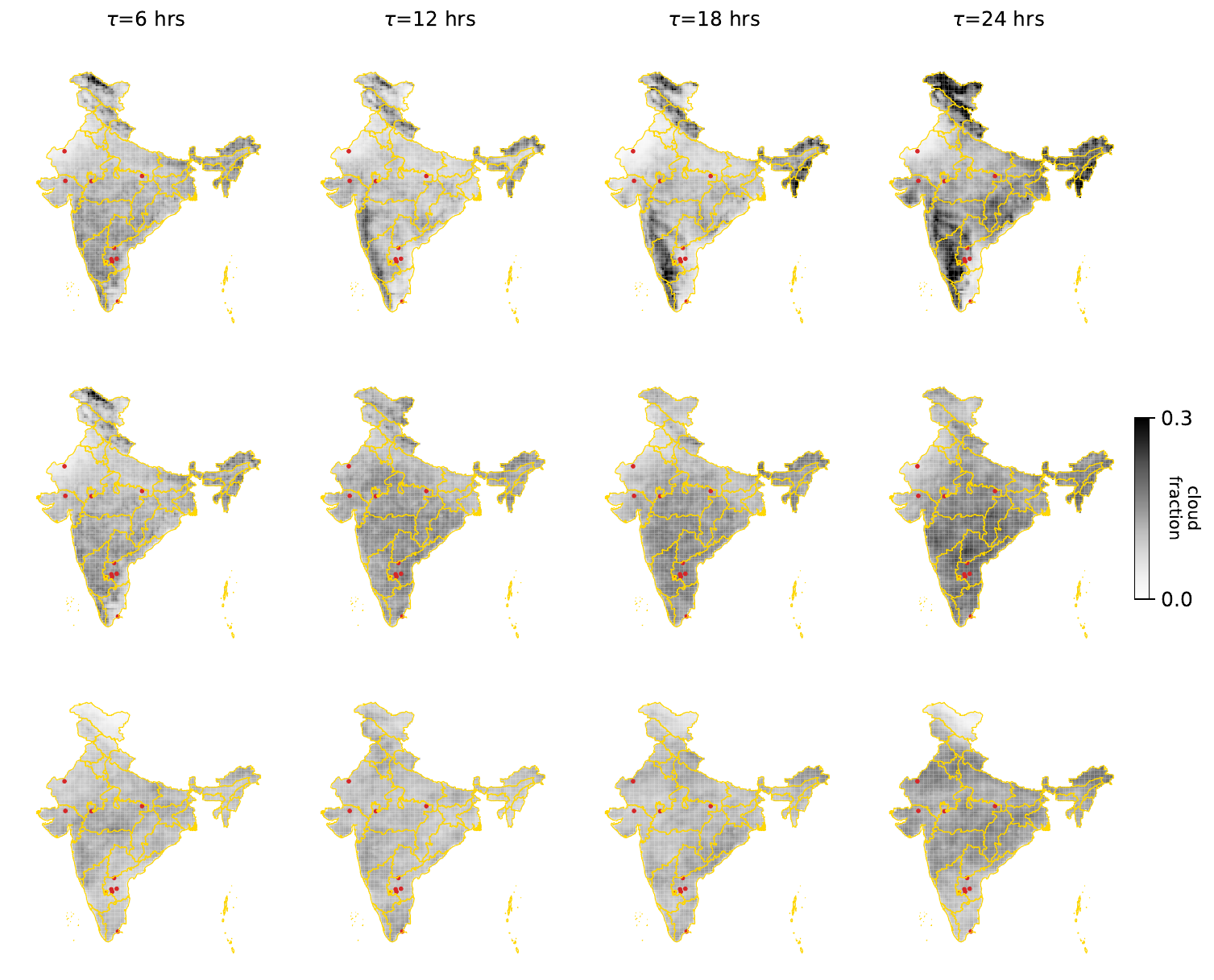}
   \caption{\small \textbf{Cloud cover validation reveals scale-dependent AI model performance across atmospheric levels.} Monsoon-averaged cloud cover differences relative to ERA5 for low-level (top row), medium-level (middle row), and high-level (bottom row) clouds across forecast lead times $\tau = 6$, 12, 18, and 24 hours. Color scales indicate cloud cover fraction differences, with distinct spatial patterns emerging across atmospheric levels. Red dots indicate locations of the 10 largest solar plants in India.}
   \label{fig:clouds_era5}
\end{figure}

We divide the errors in AIFS cloud predictions into low-, medium-, and high-cloud cover for different times of the day. Since INSAT only images cloud top properties, it is not possible to divide the retrieved cloud cover into low, medium, and high components with sufficient accuracy. So, we compute these errors w.r.t. ERA5. Comparison with ERA5 reveals substantially higher errors in predictions of low- and medium-cloud cover than predictions of high-cloud cover (Figure \ref{fig:clouds_era5} last row vs other rows). For low-cloud cover (Figure \ref{fig:clouds_era5} first row), the errors uniformly increase with lead times over the Himalayan (and foothills) regions of Northern India, Western Ghats region of Southwest India, and the high precipitation regions of Northeast India. As shown by the overlaying red dots, the Southwest region is home to multiple solar power plants, suggesting a strong impact of low-cloud nowcasting errors on solar generation estimation. The error increase (with lead time) is less dramatic for the mid cloud cover. For a 6-hour lead time, these errors are highest over the Himalayan regions in Ladakh (highly correlated with the low cloud errors), but uniformly spread over the other regions for longer lead times.

Our validation results highlight the importance of topography and precipitation in dictating cloud prediction errors, while also indicating the need for better satellite products that allow a direct model-to-observations comparison of cloud cover on a large scale.

\subsection*{Cyclone Track Prediction for Deterministic and Ensemble AI Models}

Modern weather forecasting heavily relies on ensembles of weather model initializations. Having an ensemble of model instances allows sweeping a broader probabilistic space of spatio-temporal evolution of the earth system and potentially compensates for measurement and data assimilation errors during model initialization. ECMWF's IFS, for instance, initializes a total of 50 ensemble members to forecast weather. Multiple ensembles can be critical, especially for computing cyclone trajectories. As a cyclone intensifies and approaches landfall, its trajectory can be influenced by a multitude of factors, including ocean heat flux and wind shear, none of which are globally well-constrained in weather models. Having an ensemble of initializations allows computing multiple possible trajectories of a cyclone and thus helps to constrain uncertainties in disaster management and decision-making.

\begin{figure}
   \centering
   \includegraphics[width=\linewidth]{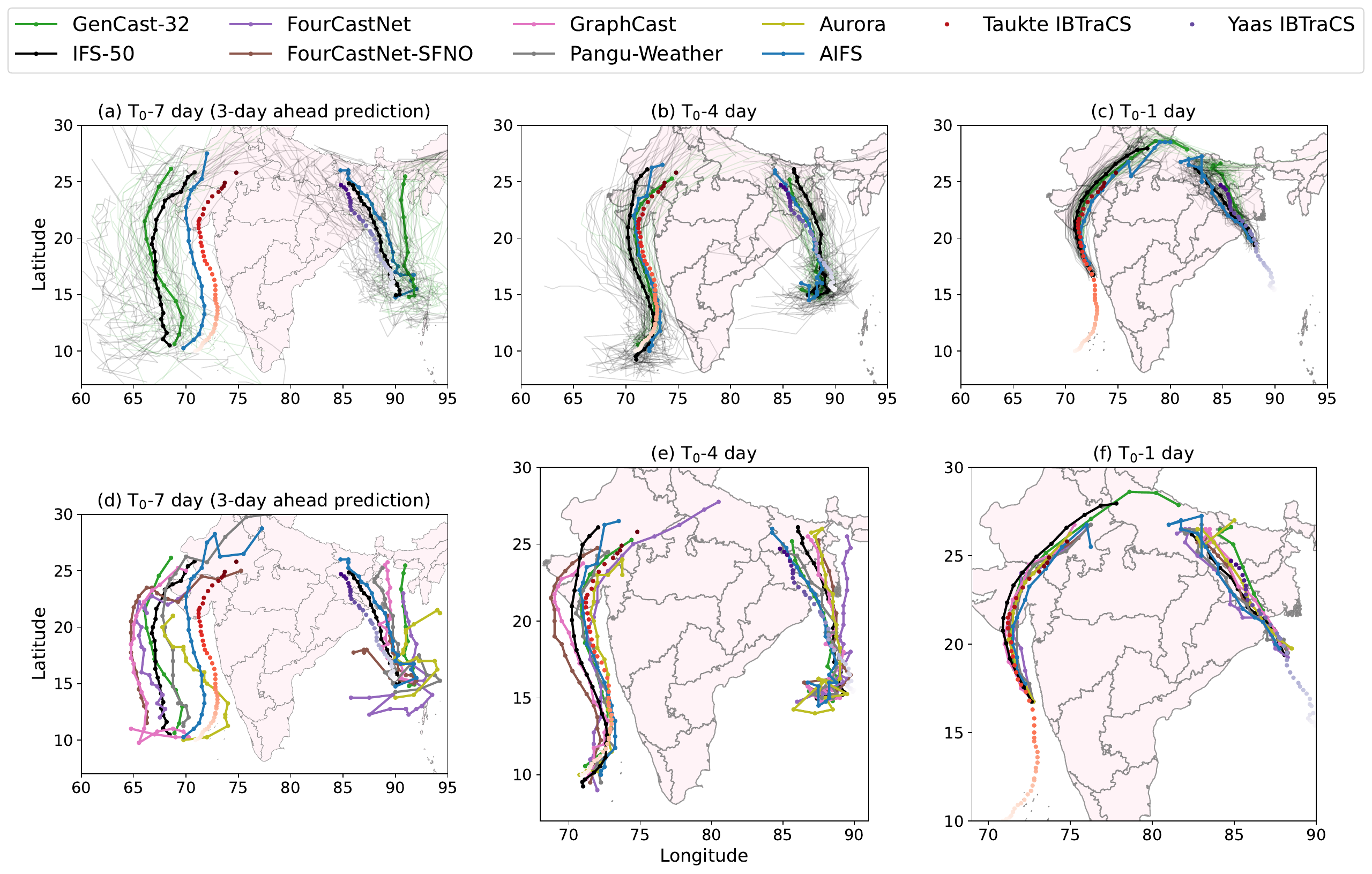}
   \caption{\small \textbf{Tropical cyclone track prediction reveals complementary strengths of ensemble and deterministic AI approaches.} Predicted trajectories for Cyclones Tauktae (Arabian Sea) and Yaas (Bay of Bengal) during May 2021, comparing GenCast-32 ensemble (green), IFS-50 ensemble (black), and six deterministic AI models (colored lines) against IBTrACS observations (red/blue dots). Panels show forecasts initialized at (a,d) T$_0$-7 day (3-day ahead prediction), (b,e) T$_0$-4 day, and (c,f) T$_0$-1 day lead times.} 
   \label{fig:cyclone_trajectories}
\end{figure}

Among all the AIWP models that can support ensembles (GenCast, Arches, and Stormer), GenCast does so at the highest spatial resolution (of 25 km). We compares GenCast ensemble's capability in simulating cyclone trajectories during the Monsoon season with the state-of-the-art ECMWF's IFS Ensembles. We considered two severe cyclones: Tauktae and Yaas, which developed over the Arabian Sea and Bay of Bengal, respectively, and battered the Indian subcontinent one week apart during Monsoon onset in 2021. A simple criterion of maximum relative vorticity at 850 hPa was used to track the cyclones' evolution. Cyclone trajectories were computed for all the deterministic AIWP model runs, the 32-member GenCast ensemble, and compared with the 50-member IFS ensemble and best-track observation data from IBTRaCS \cite{Knapp.etal2010, Gahtan.etal2024}. IBTRaCS data for Tauktae and Yaas is only available four days before landfall.

For three-day ahead forecasts made seven days before landfall, we note a substantial spread between the models and IBTRaCS trajectory (Figure \ref{fig:cyclone_trajectories}a). For Tauktae over the Arabian Sea (west), IFS ensemble presents a larger cone of uncertainty than GenCast. Interestingly, the trajectory from the deterministic AIFS (blue) is closer to IBTrACS's path than the ensemble mean trajectories from both IFS and GenCast. 

For Yaas over the Bay of Bengal (east), GenCast provides a much different trajectory for the cyclone than both IFS Ensemble and AIFS. In fact, for both the cyclones, seven-day ahead forecasts from GenCast suggest that the cyclone will not make landfall over India at all. A similar spread in a seven-day ahead forecast is seen across multiple deterministic models (Figure \ref{fig:cyclone_trajectories}d). Most models detect the cyclone, but their predicted trajectories are up to 10 degrees apart in latitude. The uncertainty is notably reduced for four-day forecasts (Figure \ref{fig:cyclone_trajectories}b,e). At this stage, almost all deterministic models and (most) ensemble members predict cyclone Tauktae's and Yaas's landfall over the Indian states of Gujarat and Odisha, respectively. For four-day ahead forecasts, both AIFS and the 32-member GenCast provide a much better prediction of the Cyclone trajectory than the 50-member IFS. Although there is a reduction in trajectory uncertainty one day before landfall, the models exhibit key differences in cyclone evolution over land. This is more readily seen over the Bay of Bengal. Here, firstly, there is a clear distinction between GenCast's and IFS Ensemble's evolution, with GenCast's trajectories substantially eastward of IFS's trajectories. Secondly, a significant spread is also obtained in cyclone evolution over land among various deterministic models (Figure \ref{fig:cyclone_trajectories}c,f).

Throughout the cyclones' evolution, IFS's ensemble members exhibit a greater spatial variability than GenCast's ensemble members. 
Part of the weak spread may be connected to a weak ensemble spread during initialization. The surface EKE spread is similar for both probabilistic models (Figure \ref{fig:ensemble_spectrum}a). However, while the ensemble spread remains almost fixed for different lead times in GenCast, the ensemble spread at IFS grows rapidly with lead time for IFS Ensemble, even though the planetary scale perturbations in IFS are two to three orders of magnitude weaker in IFS (Figure \ref{fig:ensemble_spectrum}b).

\subsection*{Implications for hyperlocal high-frequency predictions}

\begin{figure}
    \centering
    \includegraphics[width=0.5\linewidth]{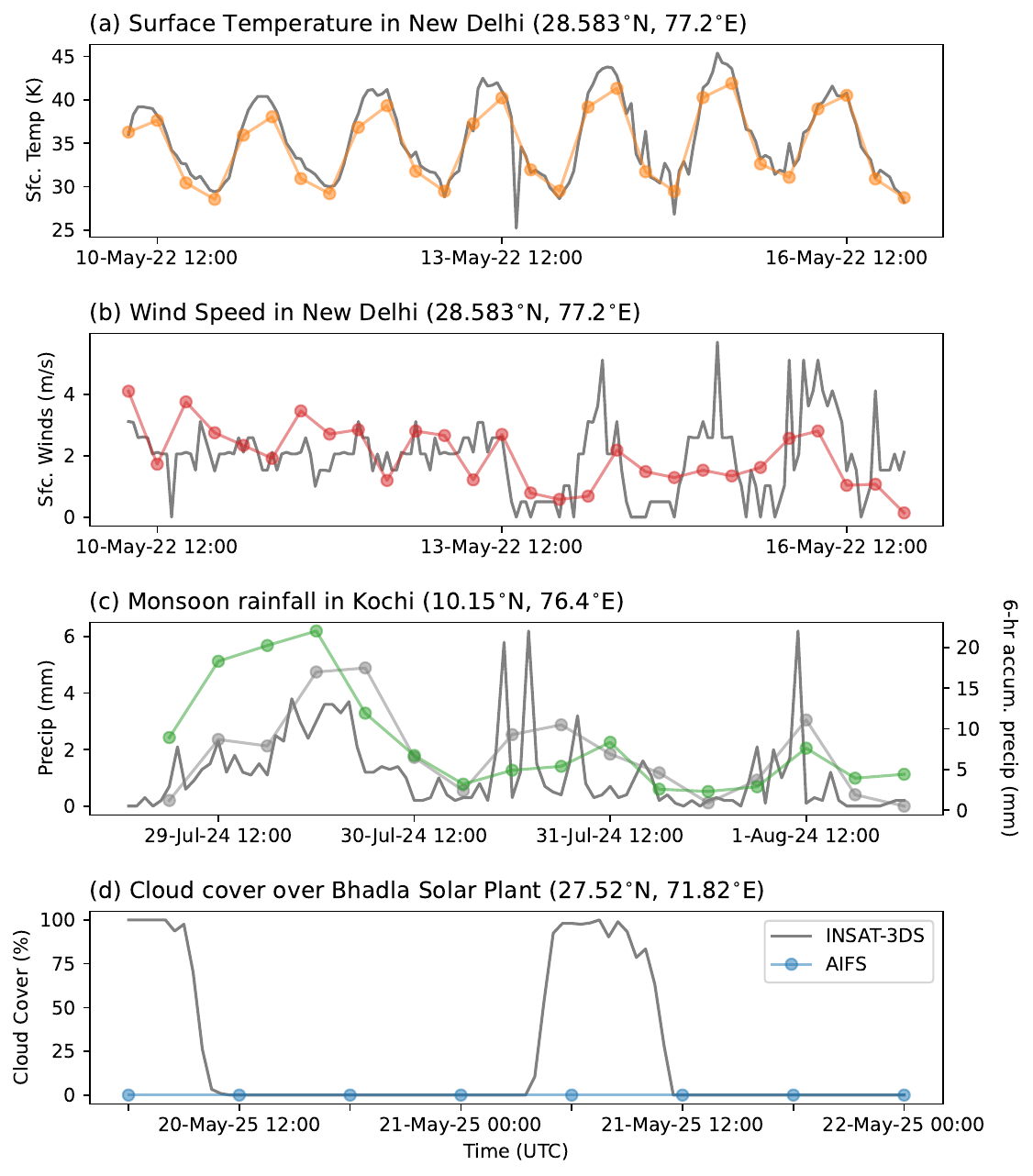}
    \caption{Hyperlocal predictions over select points: (a) temperature and (b) wind speeds respectively in New Delhi during the May 2022 heatwave, (c) extreme Monsoon precipitation in Kochi around July 2024, (d) cloud cover over Bhadla Solar Park in May 2025.}
    \label{fig:hyperlocal_errors}
\end{figure}

The presented analysis underscores both the strengths and limitations of AIWP models in emulating global weather dynamics, cloud and precipitation forecasting, and ensemble modeling of cyclones, upon being trained on ERA5 reanalysis at 25 km spatial resolution. 

At 25 km and 6-hourly space-time resolution, AIWP models miss major regional and high-frequency variability that could be critical to commercial operations (like optimizing day-ahead energy procurement for electric grids, optimizing delivery routes during storms and extreme precipitation, and predicting renewable energy generation by solar and wind farms). For instance, 6-hourly forecasts by AIFS fail to resolve the peak afternoon temperatures during the May 2022 heat wave over Delhi, underpredicting the maximum temperatures by 5 K (Figure \ref{fig:hyperlocal_errors}a). Even as the gross structure of the predicted surface temperatures is similar to the observed surface temperature, the predicted surface wind speeds are completely different from the observed wind speeds (Figure \ref{fig:hyperlocal_errors}b). Similarly, 6-hourly accumulated precipitation forecasts from AIWP models obscure the high-frequency precipitation extremes frequently observed across India during the summer monsoon, such as those shown over Kochi in Kerala in Figure \ref{fig:hyperlocal_errors}c. Relying on cloud cover prediction from AIWP models can lead to substantial errors in surface short-wave radiation (gross horizontal irradiance) estimates for solar energy generation (Figure \ref{fig:hyperlocal_errors}d). These findings motivate research towards improving the reliability and usability of AIWP models for operational applications and around high-impact weather events (\cite{Davidson.Millstein2022, Suri.etal2025}).

\section*{Discussion}

The systematic evaluation of state-of-the-art AIWP models against observations presented here provides a solid grounding to understand both their strengths and limitations. The models are trained on ERA5 and other model-derived global datasets, yet their value and usability lies in their capability to match ground-based, \emph{in situ}, or remote-sensing observations. 

Our results show that current AIWP models demonstrate considerable predictive skill against ERA5 reanalysis, and in some instances, HRES prediction during Monsoon. These models also perform reasonably well against ground-based precipitation observations from rain gauges during different stages of the Southeast Asian Monsoon, and cloud cover measurements from geostationary satellites centered over India. Most of these models also preserve global statistics by reproducing plausible planetary- and synoptic-scale temperature, kinetic energy, and humidity spectra at the surface and in the free atmosphere, indicating that even though such physical constraints were not encoded into the models, they replicate the multi-scale energy transfers in the atmosphere to an impressive degree.

After testing on multiple meteorological variables and metrics, we conclude that ECMWF's AIFS model outshines most other deterministic AIWP models both in terms of accuracy and usability. AIFS consistently projects the smallest errors overall against ground-based observations from weather stations. Even though GenCast maintains the most accurate spectra, more accurate than other models, AIFS consistently produces spectra that closely match ERA5 for various lead times, with some underestimation of power in the mesoscales. AIFS also does not exhibit power accumulation around grid scales, as other models do. Among the three models that output accumulated precipitation, AIFS outperforms GenCast and GraphCast for state-level precipitation prediction for 2024 (but not 2022). The rainfall distributions also reveal that GraphCast substantially underestimates extreme rainfall (distribution tails) more than other models, and distributions from AIFS appear to be more consistent with both IMD ground observations and ERA5. Additionally, AIFS is the only AIWP model that provides cloud cover predictions. The cloud cover from AIFS shows strong agreement with both instantaneous and time-averaged cloud satellite observations during the Monsoon season over India. Finally, the deterministic AIFS emulated the most reliable cyclone trajectories seven days ahead. For both Cyclones Tauktae and Yaas, these trajectories were substantially better than the 32-member GenCast, and for Tauktae, better than the 50-member IFS. Even for the four-day ahead forecast, both AIFS and GenCast provide a more accurate trajectory prediction than IFS.

\begin{figure}
    \centering
    \includegraphics[width=1\linewidth]{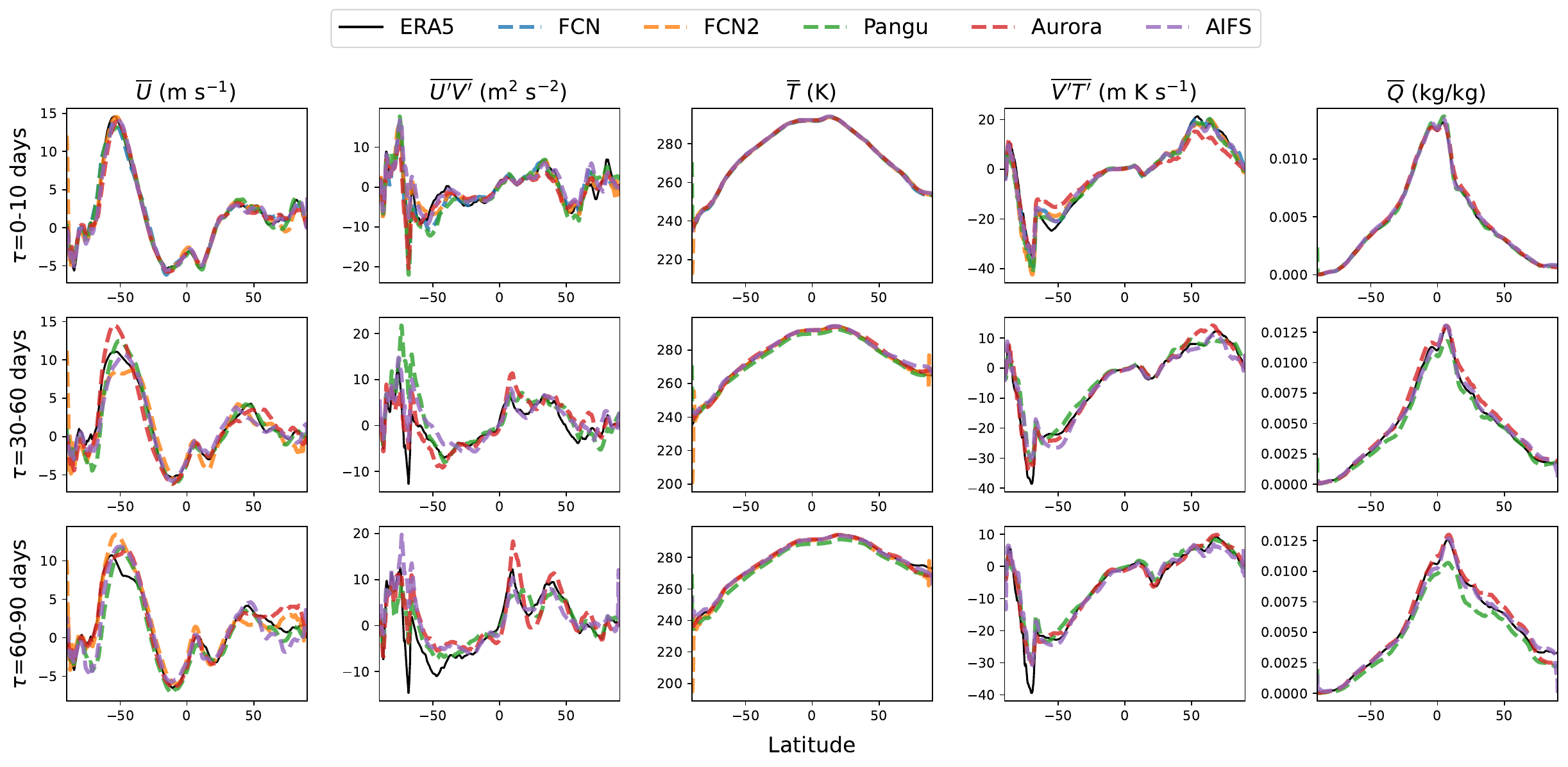}
    \caption{Zonal means and zonal eddy statistics across models for subseasonal-to-seasonal timescales at 850 hPa.}
    \label{fig:s2s_structures}
\end{figure}

{\bf Subseasonal-to-seasonal predictions:} All the models considered in the study were trained to produce up to 15-day ahead weather forecasts through auto-regressive rollouts from initial conditions. However, we also stress-tested these models by running them over subseasonal-to-seasonal (S2S) timescales of up to 90 days (initialized at 15 April 2024). Due to high memory considerations, we were unable to run Google DeepMind's GraphCast and GenCast for S2S timescales. All models, except FourCastNet, produced stable rollouts up to 90 days. When tested on key zonal mean statistics at 850 hPa, like the zonal mean zonal wind, temperature, and specific humidity structure, and eddy momentum and heat fluxes, the models match well with ERA5 for 0-10 day forecasts (Figure \ref{fig:s2s_structures}, top row). Even for higher lead times of 30-60 days or 60-90 days, the models continue to produce plausible temperature, humidity, and heat flux distributions. Most deviations from ERA5 over longer timescales are visible for the zonal mean structure and the meridional momentum fluxes. Some models develop a stronger midlatitude jet in the southern hemisphere, while some models generate another jet in the northern polar regions. These anomalies are accompanied by notable deviations in the eddy flux transport. All these features are also evident at 500 hPa. While a more detailed analysis of S2S prediction in these models is beyond the scope of this work, initial analysis sheds positive light on the capabilities of these models to maintain the general circulation of the troposphere beyond lead times that they were trained on. 

{\bf Key limitations:} Our analysis also identifies some critical limitations of current AIWP models. Working on these limitations can potentially improve the physical reliability of these models and also make them more suitable for commercial and operational applications and disaster preparedness.
\begin{itemize}
    \item {\bf Limited output variables:} Most AI models output a limited set of variables --- surface and atmospheric wind vector field, skin and atmospheric temperature, specific humidity, geopotential and mean sea-level pressure. Only 3-4 AI models output precipitation but at a 6-hour or 12-hour frequency, and only one model outputs cloud cover. Very few models output dew point temperature, and none of the models output GHI (gross horizontal irradiance), albedo, surface humidity, and actual surface pressure (as opposed to mean sea level pressure). 

    \item {\bf Underestimated mesoscale power and grid-scale power accumulation:} Our analysis revealed a systematic underestimation of power at mesoscales among all models. GenCast was a notable exception and provided a spectrum that strongly agrees with ERA5. For most models, including GenCast, however, we noted an accumulation of power near the grid-scale, indicating that the models might be generating noisy fine-scale predictions.

    \item {\bf Limited extreme weather predictability:} All of AIFS, GenCast, and GraphCast underestimate extreme rainfall during the Southeast Asian Monsoon in their day-ahead forecast. In addition, there is a notable decrease in predictive skill, especially for extreme rainfall events, for higher lead times. Moreover, the model predictions are closer to ERA5 than they are to ground-based observations. A majority of extreme rainfall variability occurs over sub-hourly or sub-6-hourly timescales. Such activity is usually recorded in weather stations (which are sparse) but is barely resolved in 6-hourly AI weather forecasts. Similarly, substantial errors in cyclone track prediction (and hence associated along-track rainfall) are noted among both deterministic and probabilistic AI models, even for four-day-ahead forecasts.

    \item {\bf High-frequency and local variability:} A lot of hyperlocal (city-level) and temporal variability is also masked by the relatively sparse (25 km) AIWP model resolution. Grid-scale error accumulation (as informed by the spectrum) and 6-hourly predictions also obscure the evolution of clouds, which have typical evolution timescales of tens of kilometers and a couple of hours. Some of these limitations can be alleviated using regional downscaling, but biases remain in such approaches, and the unavailability of high-frequency observational data makes temporal downscaling challenging.

    \item {\bf Training and Evaluation on multiple datasets and observations:} The AIWP models produce the highest errors against ground-based observations, especially wind magnitudes, suggesting that future AIWP models should not just be trained and tested on ERA5 or other global reanalyses, but also over a blend of multi-modal observations. Aurora tries to partially bridge this gap by training on multiple modeling outputs, ranging from ECMWF's HRES and ERA5 to CMIP6 output. However, key issues remain. For instance, despite training on multiple datasets, the foundation model Aurora does not offer the best performance. Notable deviations from ERA5's spectrum at both small- and planetary-scales for $\tau=7$ days and $\tau=10$ days raise issues regarding physical consistency in the model. More advanced training strategies or modeling architectures might be required to ensure a physically consistent inclusion of multiple modalities into the AI models.  
\end{itemize}
Addressing and solving these limitations will ensure a wide usability of the next-generation AI-based weather prediction models, unleashing their true value for meteorological analysis, operational weather prediction, commercial applications, and disaster preparedness.

\section*{Materials and Methods}


\subsubsection*{FourCastNet and FourCastNet-SFNO}
FourCastNet \cite{Pathak.etal2022} (FCN in short), developed at NVIDIA, represents pioneering efforts in transformer-based weather forecasting. FourCastNet-SFNO (or FourCastNetv2) represents an update to the initial FourCastNet model through the use of Spherical Fourier Neural Operators \cite{Bonev.etal2023}. Both FourCastNet and FourCastNet-SFNO were trained on 6-hourly ERA5 reanalysis dataset from 1979 to 2015. Years 2016 and 2017 were used for validation. Years 2018 and onwards were used for testing. In the AI-models implementation of FourCastNet, the model predicts, 2metre surface temperature (T2M), 10metre zonal and meridional winds (U10, V10), and mean sea level pressure (MSLP) at the surface, and winds (U,V), temperature (T), geopotential (Z), and relative humidity (RH) in the free atmosphere at a 0.25$^{\circ} \times$ 0.25$^{\circ}$ spatial, and 6-hourly temporal resolution.

\subsubsection*{Pangu-Weather}
Developed by Huawei Technologies Co., Pangu-Weather \cite{Bi.etal2023} uses a 3D-Earth Specific Transformer trained on ERA5 reanalysis from years 1979 to 2017, validated on 2019 ERA5 data, and tested on 2018 ERA5 data. The model predicts, T2M, U10, V10, and MSLP at the surface, and U V, T, Z, and specific humidity (Q) on 13 pressure levels in the free atmosphere at 0.25$^{\circ} \times$ 0.25$^{\circ}$ spatial resolution. Pangu-Weather has both a 6-hourly forecast version and an hourly forecast version. In this study, we use the 6-hourly version only. The 13 pressure levels are: 50, 100, 150, 200, 250, 300, 400, 500, 600, 700, 850, 925, and 1000 hPa.

\subsubsection*{GraphCast and GenCast}
GraphCast \cite{Lam.etal2023} and GenCast \cite{Price.etal2025} are leading AI weather prediction models developed by Google DeepMind. GraphCast uses a graph neural network-based attention network, while GenCast uses a conditional diffusion transformer to generate ensembles of forecasts. GraphCast was trained on ERA5 reanalysis from 1979 to 2018 and fine-tuned on ECMWF HRES from 2016 to 2021, and GenCast was trained on ERA5 reanalysis from 1979-2018. Both models generate outputs at 0.25$^{\circ} \times$ 0.25$^{\circ}$ spatial resolution. GraphCast provides predictions every 6 hours, while GenCast provides predictions every 12 hours. Both models output the same set of output variables: T2M, U10, V10, MSLP, sea surface temperature (SST), and total precipitation (TP) at the surface, and U, V, T, Z, Q, and vertical velocity (W) on 13 pressure levels in the atmosphere. We run GenCast both in the deterministic and ensemble configuration. For WeatherBench-like error analysis, we focus on the deterministic GenCast runs, but for cyclone trajectory comparison, we run a 32-member GenCast ensemble.

\subsubsection*{Aurora Foundation Model}
Aurora \cite{Bodnar.etal2025}, developed by Microsoft Research, is one of the first foundation models for weather applications. A weather foundation model is pre-trained on multiple years of weather forecasting data. However, unlike other AI weather models, a foundation model's architecture allows it to be fine-tuned to perform other weather-related tasks as well. In the case of Aurora, along with weather forecasting, it has been fine-tuned to perform air quality, ocean wave modeling, and hurricane prediction tasks as well. Aurora uses a multiscale 3D Swin Transformer U-Net backbone and allows producing weather forecasts at both a 0.25$^{\circ} \times$ 0.25$^{\circ}$ and 0.1$^{\circ} \times$ 0.1$^{\circ}$ spatial resolution. For consistency with other models, we use the 0.25$^{\circ} \times$ 0.25$^{\circ}$ weather forecasting checkpoints. Aurora was trained on a multitude of datasets, including ERA5 reanalysis, HRES operational forecasts, IFS ensemble forecasts GFS operational forecasts, GEFS ensemble reforecasts, CMIP6 climate simulations, MERRA-2 atmospheric reanalysis, as well as CAMS forecasts, analysis and reanalysis data. Aurora outputs T2M, U10, V10, MSLP at the surface, and T, U, V, Q, and Z on 13 pressure levels in the atmosphere.

\subsubsection*{AIFS}
AIFS \cite{Lang.etal2024} is ECMWF's deterministic Data-driven forecasting system based on graph neural network encoder and decoder. The model both inputs and outputs meteorological variables on the N320 reduced Gaussian grid. Similar to GraphCast, the model is first trained on 1979-2018 ERA5 reanalysis data, and then fine-tuned on ECMWF's operational NWP analyses data. Similar to other models, the model generates output on the N320 Gaussian grid and 6-hourly time step. For this study, the output was regridded from the N320 grid to a 0.25$^{\circ} \times$ 0.25$^{\circ}$ uniform lat-lon grid. In addition outputting T2M, U10, V10, MSLP at the surface, and T, U, V, W, Q, and Z on 13 pressure levels, AIFS also outputs total column water, surface pressure, total precipitation, convective precipitation, and low-, medium-, and high-cloud cover.

\subsection*{Traditional Numerical Weather Prediction Models}

For the WeatherBench-like error analysis, we compare the AIWP models against a suite of traditional models including (a) ECMWF's high-resolution deterministic HRES model and its analysis (initialization data) which provides global 9 km forecasts every 12 hours up to 10 days ahead, (b) ECMWF's 50-member IFS Ensemble model which provide probabilistic forecasts every 6 hours at an 18 km resolution up to 15 days ahead, and (c) IFS Ensemble mean forecasts, which represent the mean of the 50-ensemble IFS ensemble model predictions. We use the predictions from these models for the year 2022 provided in the WeatherBench data buckets on Google Cloud Storage.

\subsection*{Validation Data}
{\bf ERA5 reanalysis:} We test the model predictions against ECMWF's 6-hourly ERA5 renanalysis data \cite{Hersbach.etal2020} at a 0.25$^{\circ} \times$ 0.25$^{\circ}$ spatial resolution for the years 2021-2024.

{\bf MeteoStat Ground-based Weather Station Data:} We use MeteoStat Python API (\url{https://meteostat.net/en/}) to access hourly, publicly available point-based weather data from IMD's weather stations network. We use these observations to validate AI predictions of surface variables, including T2M, U10, and V10, and to assess AIFS's precipitation prediction during extreme monsoon rainfall in Kerala in Figure \ref{fig:hyperlocal_errors}. 

{\bf IMD Gridded Precipitation Data:} To compare precipitation prediction against ground-based in-situ observations, we use the daily-averaged 0.25$^{\circ} \times$ 0.25$^{\circ}$ resolution gridded rainfall data from the Indian Meteorological Department (IMD). The data is collected using up to 6995 rain gauges nationwide \cite{Pai.etal2014} and is publicly available on IMD's website. We use the data for years 2022 and 2024.

\subsubsection*{INSAT-3DS Geostationary Satellite}
To validate the total cloud cover output inferred from IFS, we use the INSAT-3DS \cite{Giri.etal2023} geostationary satellite's processed output of the cloud fraction data. Launched into orbit on 17 February 2024, INSAT-3DS is a next-generation successor to the INSAT-3DR geostationary satellite. Both INSAT-3DR and INDAT-3DS are centered at different geostationary longitudes but measure cloud top properties at the same resolution. While we couldn't find any study that validation cloud cover in INSAT-3DS, the cloud cover in INSAT-3DR has been sufficiently validated by past studies. \cite{Lima.etal2019, Gopikrishnan.etal2023, Parihar.etal2024}. To validate the cloud cover output from AIFS, we use the  0.5$^{\circ} \times$ 0.5$^{\circ}$ processed clear sky cloud fraction to infer the cloud cover. Since INSAT-3DS uses imaging in different bands across visible, infrared and water vapor channels, this could lead to a potential underestimation of the medium- and low-cloud covers in INSAT-3DS measurements.

\subsubsection*{IBTRaCS Cyclone Trajectory Data}
Cyclone trajectory predictions from the deterministic models, the 50-member IFS Ensemble model, and the 32-member GenCast ensembles, were compared against IBTRaCS best-track data \cite{Gahtan.etal2024}. The best tracks data for Cyclones Tauktae and Yaas are respectively available at \url{https://ncics.org/ibtracs/index.php?name=v04r01-2021133N10071} and \url{https://ncics.org/ibtracs/index.php?name=v04r01-2021143N15090}.

\subsubsection*{Mean Absolute Error (MAE) to quantify errors}
We use the mean absolute error to quantify errors in the WeatherBench-like error analysis across AI models, traditional models, and observations. The mean absolute error over multiple samples is computed as:

\begin{equation}
    MAE(y_p,y_t) = \frac{1}{N}\Sigma_{i=1}^N\left|y_{p,i} - y_{t,i}\right|
\end{equation}
where $y_p,i$ is the $i$-th predicted sample and $y_t,i$ is the $i$-th observational truth, respectively, and $N$ is the number of samples over which the error is calculated.

\subsubsection*{Computing EKE and Temperature Power Spectrum}
The eddy kinetic energy (EKE) spectra were computed by first decomposing the zonal and meridional wind fields into individual wave components (wavenumbers) and then computing the kinetic energy for the wavenumbers. More precisely, the integrated EKE spectrum $\mathcal{E}$ as a function of wavenumber $\kappa$ can be expressed as:

\begin{equation}
    \mathcal{E}(\kappa) = \iint_{\mathcal{D}} U(\kappa)^2 + V(\kappa)^2 \; d\lambda d\phi
\end{equation}

where $\mathcal{D}$ is the global domain, $\kappa$ is the total wavenumber, $\lambda$ is the longitude, $\phi$ is the latitude, and $U(\kappa)$ is the eddy zonal wind associated with wavenumber $\kappa$ and is obtained by first transforming the full zonal wind from physical space to spectral space, then zeroing out contributions from all wavenumbers except $\kappa$ and then transforming back to the physical space again. Similarly for $V(\kappa)$. For our 720-latitude grid, $\kappa$ ranges from 0 to 720.

The power spectrum for surface temperature is computed as:
\begin{equation}
    \mathcal{T}(\kappa) = \iint_{\kappa_x,\kappa_y} T(\kappa_x', \kappa_y')\cdot \overline{T(\kappa_x', \kappa_y')} \delta(\kappa - \sqrt{\kappa_x^2 + \kappa_y^2}) d\kappa_x' d\kappa_y'
\end{equation}
where $\kappa_x$ and $\kappa_y$ are the directional wavenumbers and the overbar denotes the complex conjugate of the spectral coefficient.

\subsubsection*{Computing Cloud Cover}
ECMWF's AIFS model predicts low-, medium-, and high-cloud cover, with each having a value between 0 and 1 and indicating the fraction of the grid cell covered by each cloud type. Since INSAT-3DS outputs the total cloud cover, we combine the individual cloud cover into the total cloud cover as:

\begin{equation}
TCC_{AIFS}  = 1. - (1 - LCC_{AIFS})\cdot(1 - MCC_{AIFS})\cdot(1 - HCC_{AIFS})
\end{equation}

where TCC, LCC, MCC and HCC, respectively are the total, low, medium, and high cloud cover as a function of latitude and longitude respectively.

\subsubsection*{Computing Cyclone Trajectories}

A simple algorithm involving tracking the maximum relative vorticity at 850 hPa was used to track the cyclone eye. First, the zonal and meridional wind fields $u$ and $v$ were used to compute the relative vorticity, i.e., the horizontal curl $\nabla_H\times(u,v)$ of the horizontal velocity. Then the latitude and longitude associated with the maximum potential vorticity were identified in a small sub-domain over the Arabian Sea and the Bay of Bengal. This was considered the starting point. Subsequently, only the 10$^{\circ} \times$ 10$^{\circ}$ box centered around the previously located vorticity maximum was used to determine the vorticity maximum at the next forecast step.

\section*{Acknowledgments}
We thank Peter Dueben and Ramalingam Saravanan for insightful discussions on operational weather forecasting and global precipitation products, respectively. We also thank Mohak Mangal for discussions on operational energy applications of weather forecasting.

\paragraph*{Funding:}
Aditi Sheshadri and Aman Gupta are supported by Schmidt Sciences, LLC, as part of the Virtual Earth
System Research Institute (VESRI). Aditi Sheshadri also acknowledges support from the National Science Foundation through Grant OAC-2004492.
\paragraph*{Author contributions:}
AG, DS, and AS conceptualized the study. AG and DS worked on the methodology. AG conducted the formal analysis, validation, and figures. AS supervised the project and provided resources for the project. All authors edited the draft.
\paragraph*{Competing interests:}
The authors have no competing interests
\paragraph*{Data and materials availability:}
All the model weights used in this study are available on HuggingFace, and all the data used to initialize the models are freely available on ECMWF's Copernicus CDS server. ERA5 reanalysis dataset is publicly available at: \url{https://cds.climate.copernicus.eu/datasets/reanalysis-era5-pressure-levels}. The model weights and code for FourCastNet, FourCastNet-SFNO, Pangu-Weather, GraphCast, and Aurora, used in this study, is available through the ECMWF AI-Models package: \url{https://github.com/ecmwf-lab/ai-models}. The Code, model weights and sample initialization files are available at: \url{https://github.com/google-deepmind/graphcast}. ECMWF's AIFS model weights are available at: \url{https://huggingface.co/ecmwf/aifs-single-1.0}.

\bibliographystyle{unsrt}
\bibliography{biblio}

\newpage

\appendix

\section{List of Models and Observations}

\begin{table}[!ht] 
	\centering
	\caption{\textbf{Models and Datasets Description.}
		All models are run at a spatial resolution of 0.25$^{\circ}$ $\times$ 0.25$^{\circ}$ (roughly 25 km). All traditional model outputs used in the study were used at a 0.25$^{\circ}$ $\times$ 0.25$^{\circ}$ resolution as well. Pangu-Weather is available at three different temporal resolutions. We use the 6-hourly model in this study.}
	\label{tab:example} 
	
	\begin{tabular}{lccl} 
		\\
	    \hline
            Observations & Source & Sampling Frequency\\
            \hline
            
            Weather Stations sfc winds and temperature & MeteoStat  & Hourly\\

            Precipitation & IMD gridded data (CITE) &  Daily averaged\\

            Cloud Cover & INSAT 3DS &  30 minutes\\

            Cyclone Trajectory & IBTrACS & Hourly\\

            \hline
            Traditional Model & Institution & Forecast step\\
            \hline
            ERA5 t=0 analysis & ECMWF  & 1h\\

            HRES t=0 analysis & ECMWF  & 1h\\

            HRES forecast & ECMWF & 12h\\

            IFS Mean forecast & ECMWF & 6h\\

            IFS ensemble forecast & ECMWF & 6h\\
            
            \hline
		AIWP Model & Institution & Forecast step\\
		\hline
		FourCastNet  & NVIDIA & 6h\\
	
        FourCastNet-SFNO & NVIDIA & 6h\\
	
        Pangu-Weather & Huawei  & 1h/6h/24h\\
        
        GraphCast & Google DeepMind & 6h\\
        
        Aurora & Microsoft & 6h\\
        
        AIFS Deterministic &    ECMWF &  6h\\
        
        GenCast Ensemble & Google DeepMind & 12h\\
		\hline
	\end{tabular}
\end{table}

\clearpage
\section{Global forecast skill}

\begin{figure}[!ht]
    \centering
    \includegraphics[width=1\linewidth]{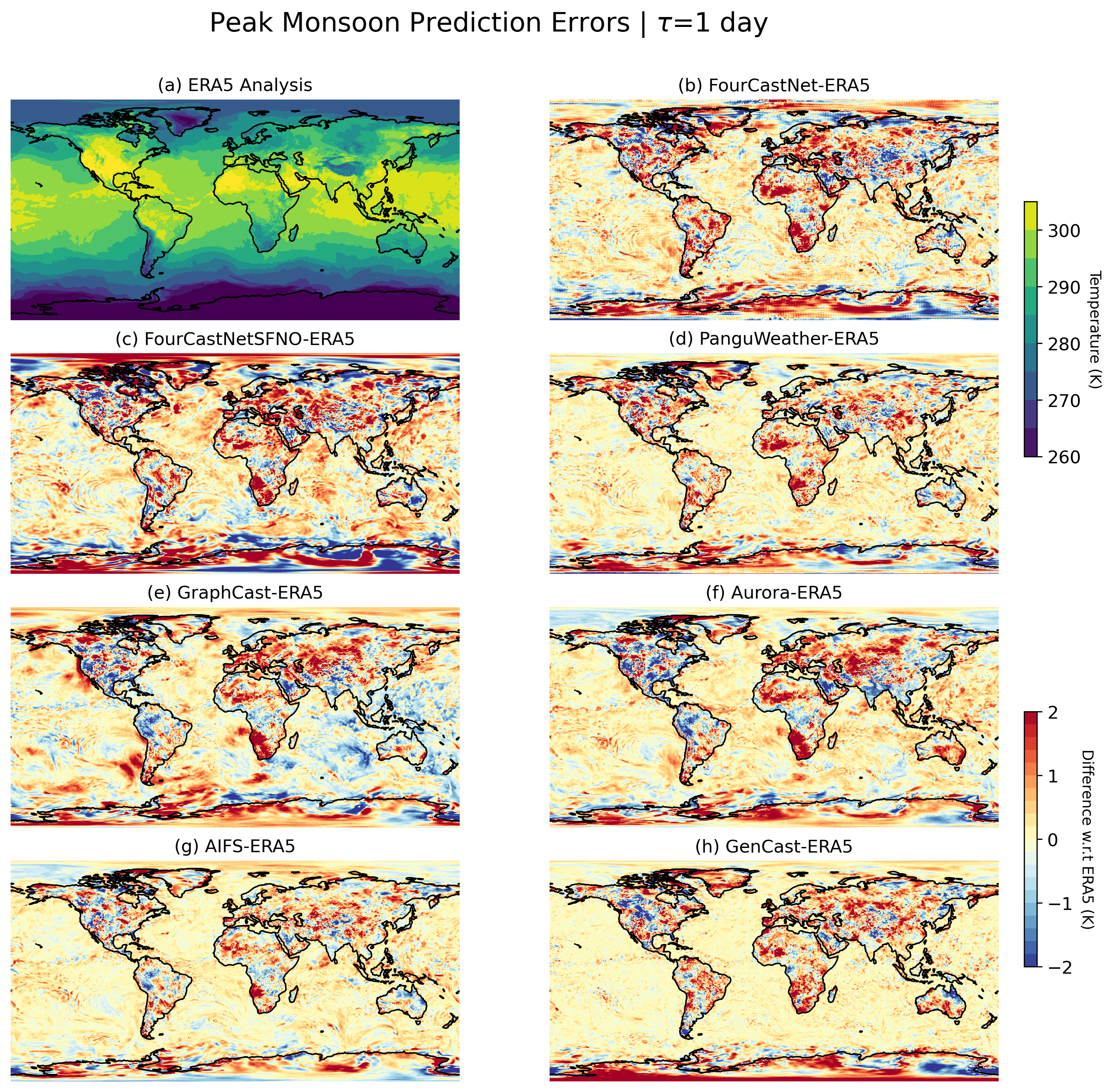}
    \caption{Global anomalies in day-ahead weather prediction across seven state-of-the-art AIWP models for models initialized on 15 July 2022 0000UTC.}
    \label{fig:global_errors}
\end{figure}

\begin{figure}
    \centering
    \includegraphics[width=\linewidth]{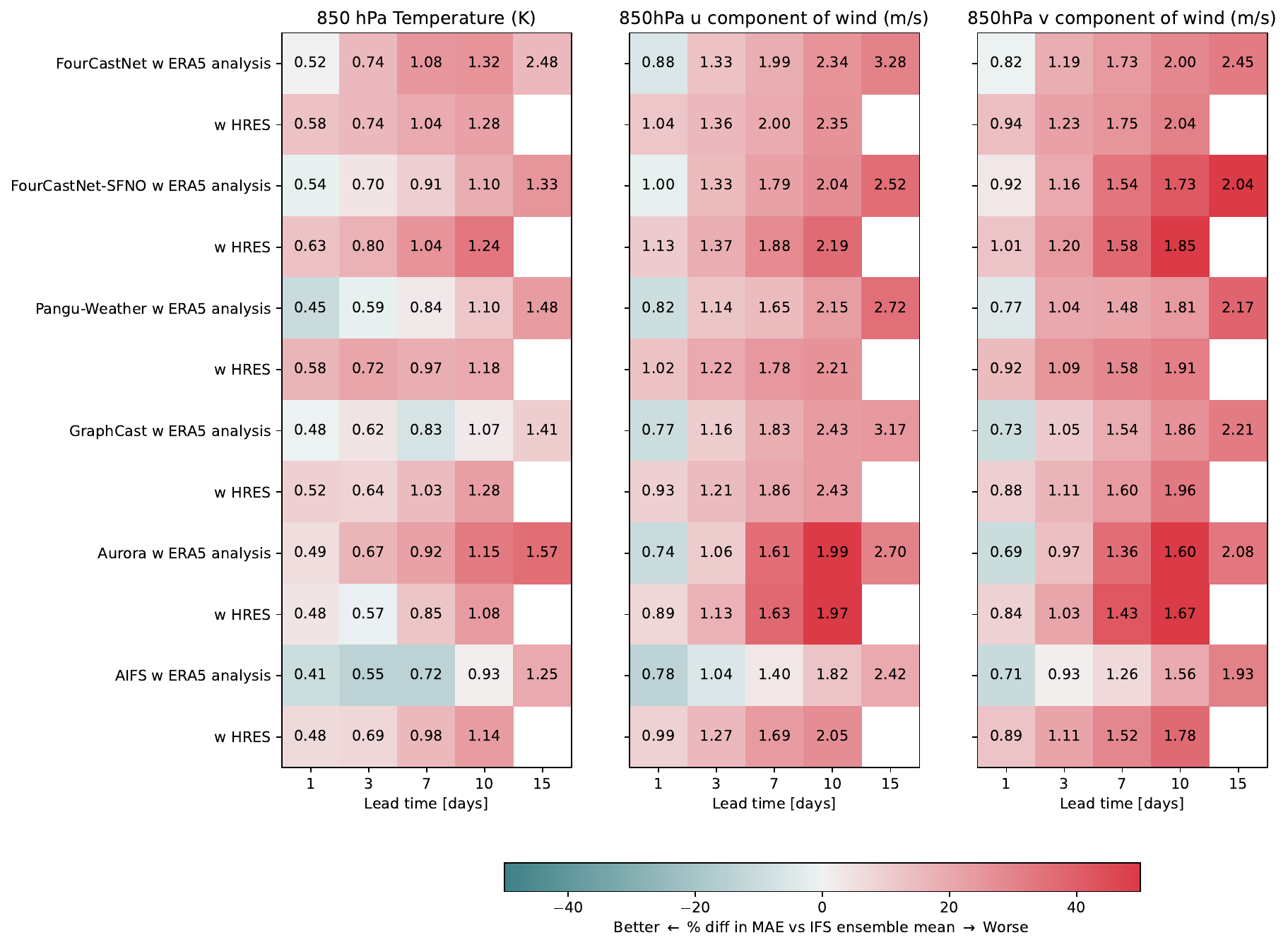}
    \caption{WeatherBench-like error plots similar to Figure \ref{fig:weatherbench_comparison} but for temperature, zonal and meridional winds at 850 hPa. Since weather station observations are only available at the surface, variables at this level are compared only against ERA5 analysis, HRES predictions, and IFS ensemble mean predictions. AIFS continues to produce the most accurate predictions for all three variables.}
    \label{fig:weatherbench850hpa}
\end{figure}

\begin{figure}
    \centering
    \includegraphics[width=\linewidth]{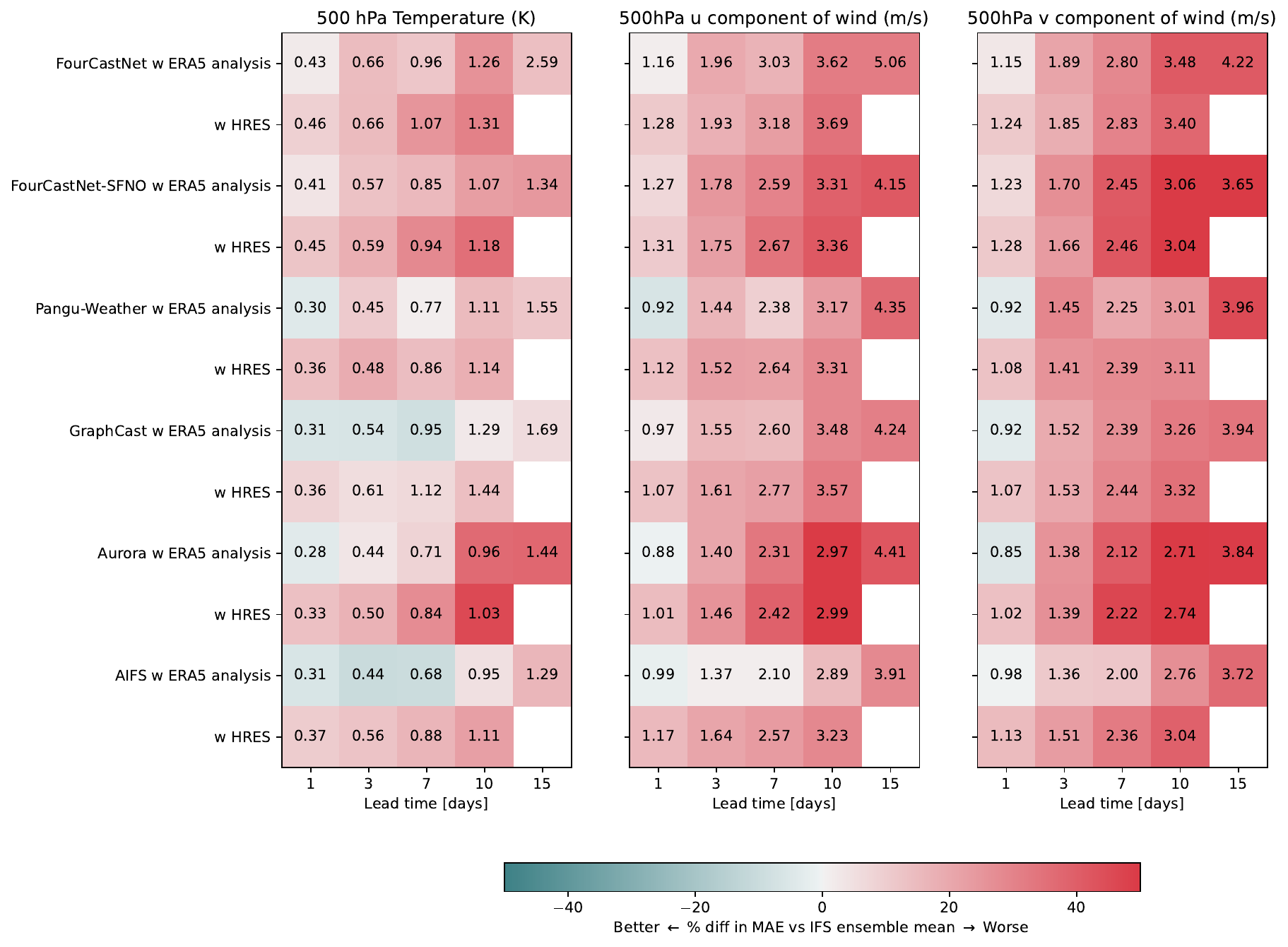}
    \caption{WeatherBench-like error plots similar to Figure \ref{fig:weatherbench_comparison} but for temperature, zonal and meridional winds at 500 hPa. Since weather station observations are only available at the surface, variables at this level are compared only against ERA5 analysis, HRES predictions, and IFS ensemble mean predictions.  AIFS continues to produce the most accurate predictions for all three variables.}
    \label{fig:weatherbench500hpa}
\end{figure}

\clearpage
\section{Global Spectrum}

\begin{figure}[!ht]
   \centering
   \includegraphics[width=\linewidth]{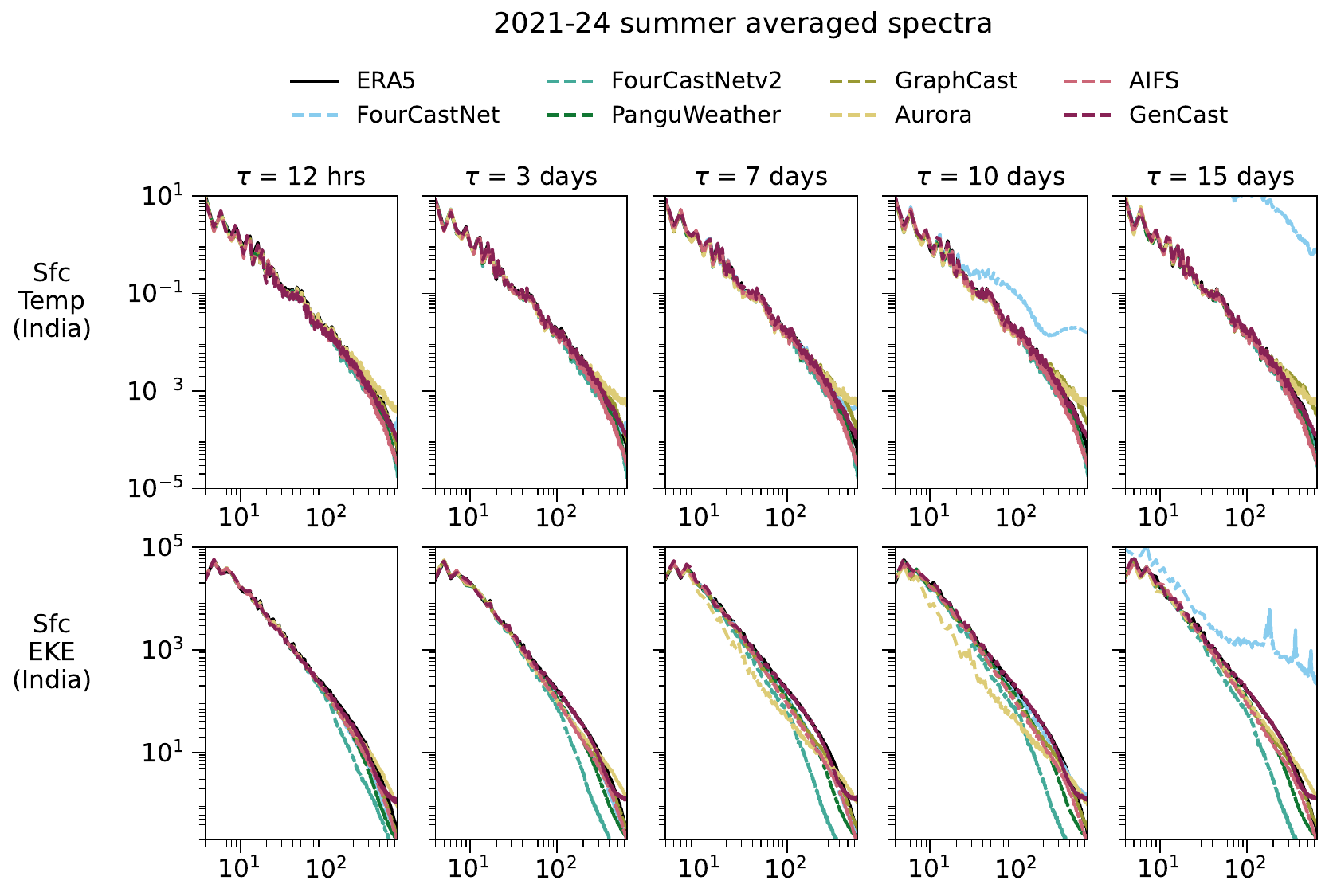}
   \caption{\small \textbf{Multi-year spectral analysis confirms systematic AI model deficiencies in energy cascade representation.} Power spectral density comparison for surface temperature (top row) and surface eddy kinetic energy (bottom row) over the Indian domain during monsoon seasons 2021--2024, averaged across May--July periods. Forecast lead times range from $\tau = 12$ hours to $\tau = 15$ days. The remarkable consistency of spectral biases across multiple years demonstrates that energy deficits at mesoscales ($\kappa \approx 10^1$--$10^2$ km$^{-1}$) represent fundamental limitations rather than seasonal artifacts. All AI models exhibit progressive spectral degradation with lead time, with characteristic energy collapse at $\tau = 15$ days where realistic mesoscale variability is artificially suppressed. This systematic spectral filtering behavior has profound implications for regional weather prediction and extreme event forecasting capabilities.}
   \label{fig:spectrum_multiyear}
\end{figure}

\clearpage
\section{Precipitation Predictions}

\begin{figure}[!ht]
   \centering
   \includegraphics[width=\linewidth]{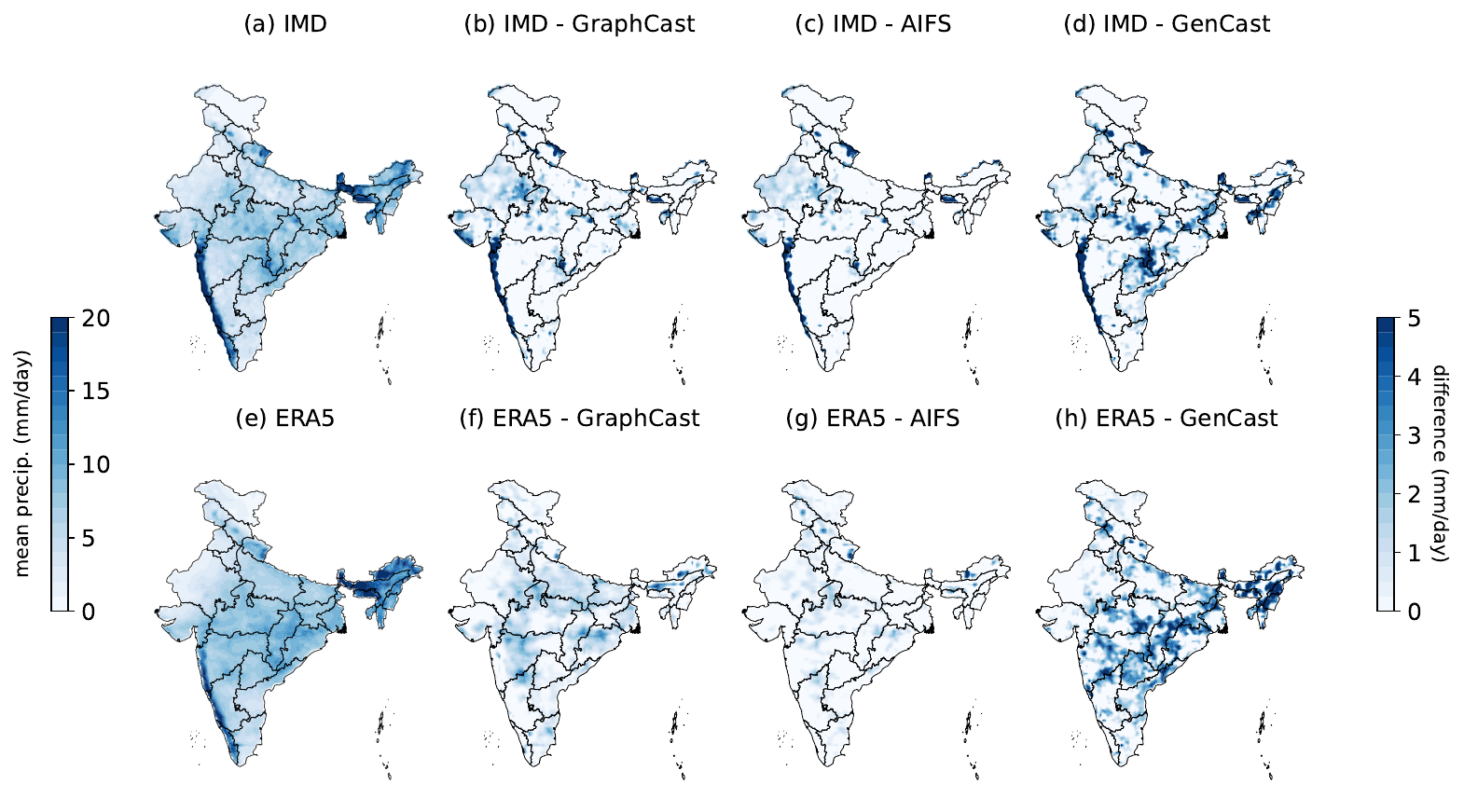}
   \caption{\small \textbf{Systematic spatial precipitation biases persist across contrasting monsoon conditions.} Mean precipitation fields and model-observation differences for the 2024 monsoon season, showing (a) IMD observations and (b-h) AI model biases relative to observations for GraphCast, AIFS, GenCast, and ERA5. Color scales indicate mean precipitation rates (mm/day) and bias magnitudes. All AI models exhibit characteristic dry biases over the Western Ghats and wet biases over central India, with coherent spatial patterns indicating systematic deficiencies in representing orographic precipitation processes. The persistence of these bias patterns across different meteorological conditions suggests fundamental limitations in monsoon precipitation physics rather than sensitivity to specific atmospheric states.}
   \label{fig:precip_2024}
\end{figure}

\begin{figure}
  \centering
  \includegraphics[width=\linewidth]{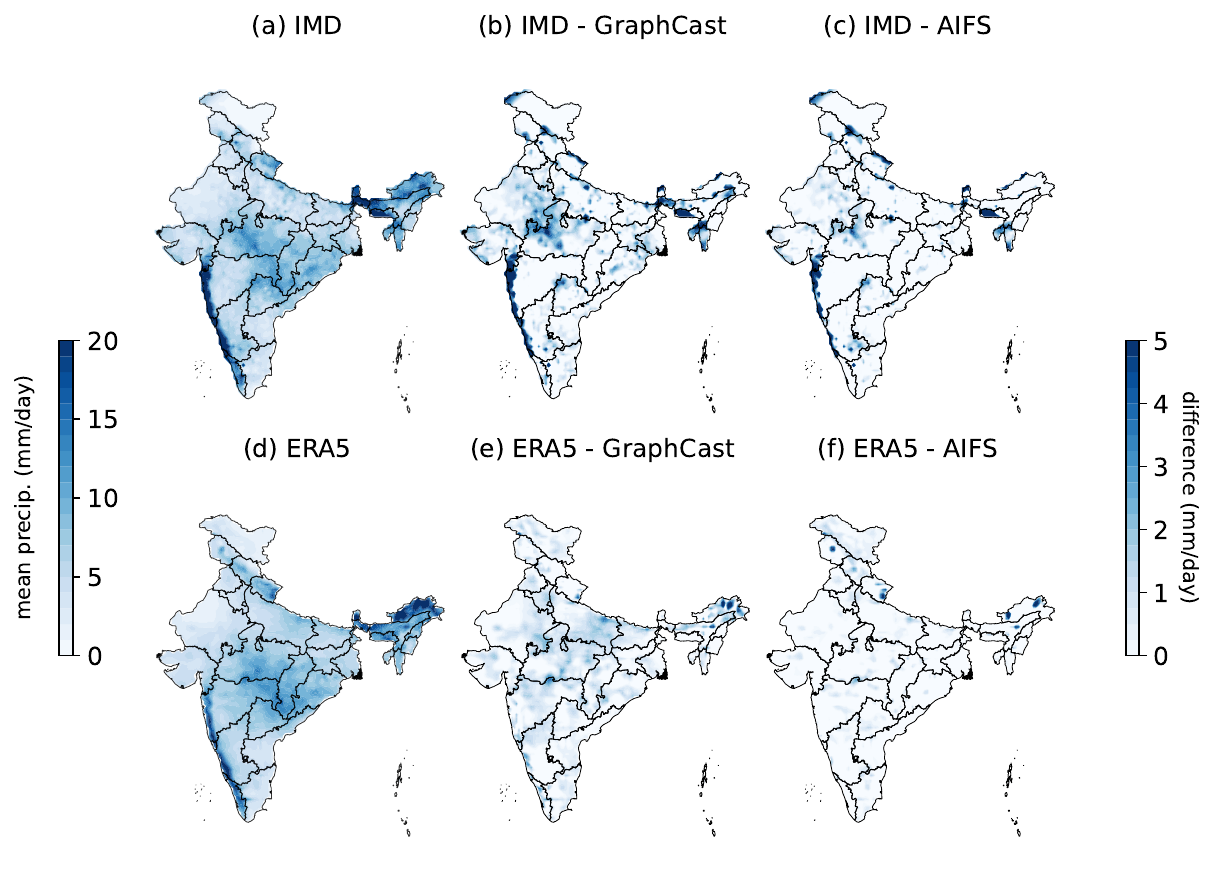}
  \caption{\small \textbf{Multi-year consistency confirms systematic nature of AI model precipitation biases.} Mean precipitation fields and model-observation differences for the 2022 monsoon season, showing (a) IMD observations and (b-f) AI model biases relative to observations for GraphCast, AIFS, and ERA5. GenCast was not available for this analysis period. Color scales indicate mean precipitation rates (mm/day) and bias magnitudes. The remarkable similarity in bias patterns between the 2022 and 2024 seasons demonstrates that observed precipitation errors represent systematic model deficiencies rather than artifacts of specific meteorological conditions. Persistent dry biases over orographic regions and wet biases over central plains indicate fundamental limitations in representing monsoon precipitation physics that transcend interannual variability in atmospheric circulation patterns.}
  \label{fig:precip_2022}
\end{figure}

\begin{figure}
  \centering
  \includegraphics[width=\linewidth]{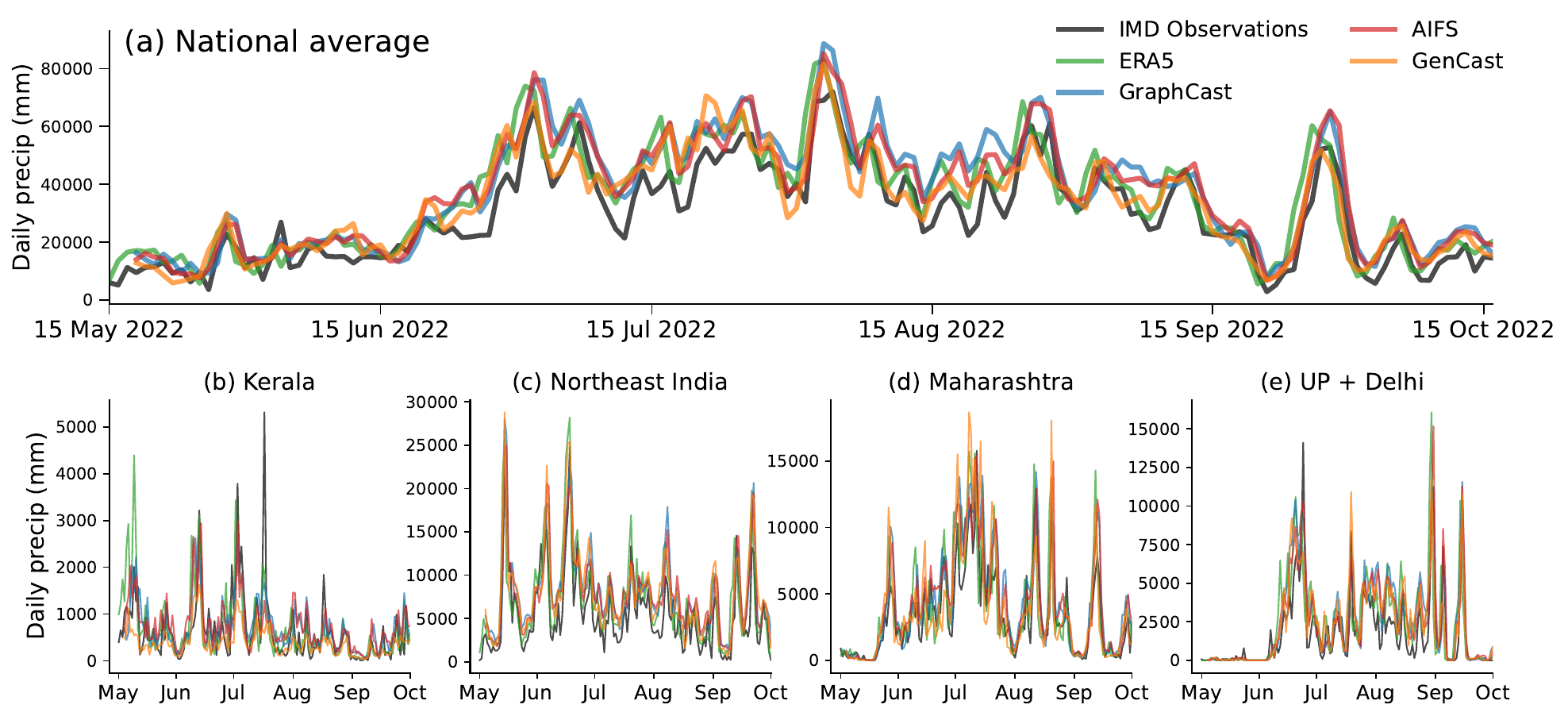}
  \caption{\small \textbf{Precipitation biases considerably increase for 3-day ahead forecasts.} Same as Figure \ref{fig:precip_timeseries} but for 3-day ahead forecasts.}
  \label{fig:precip_timeseries_lag3}
\end{figure}

\begin{figure}
  \centering
  \includegraphics[width=\linewidth]{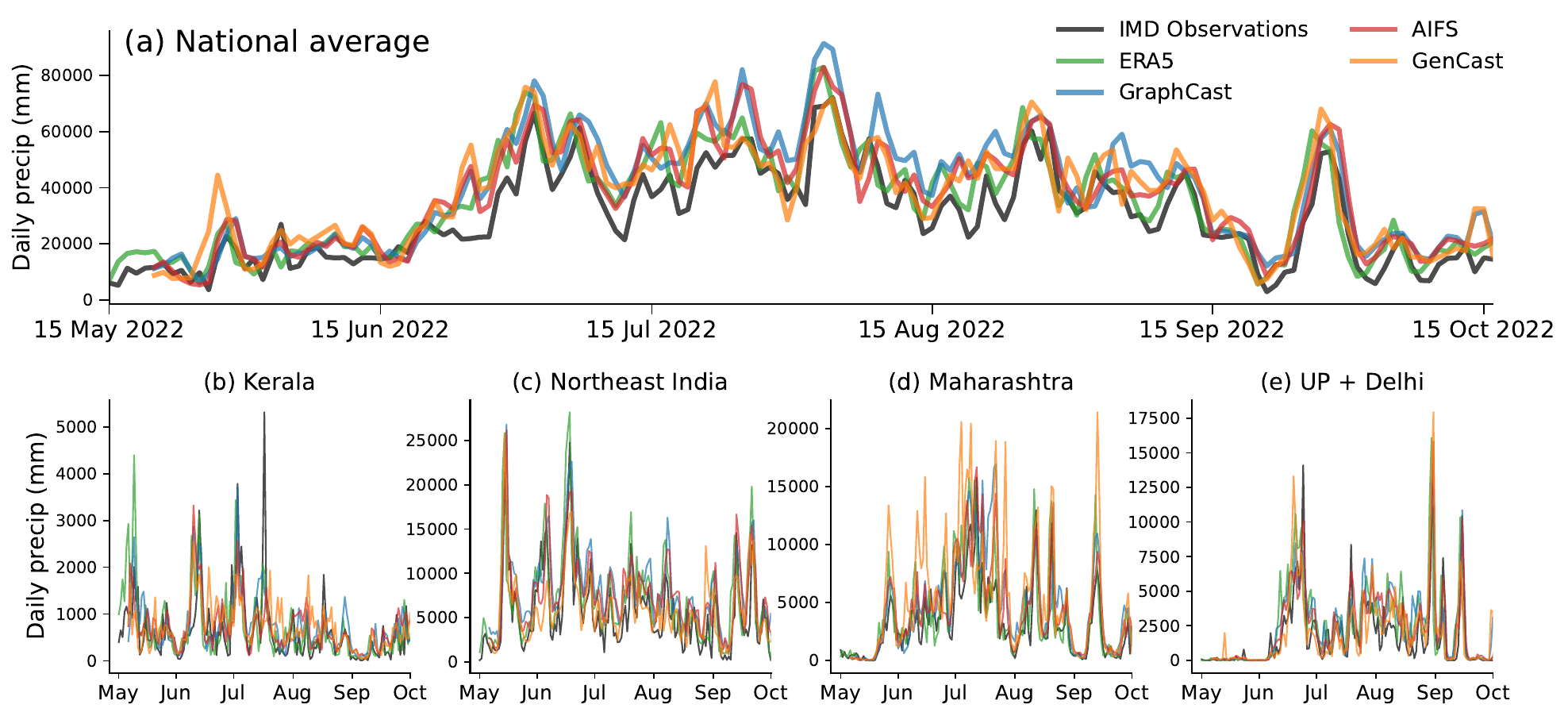}
  \caption{\small \textbf{Precipitation biases aggravate for 5-day ahead forecasts.} Same as Figure \ref{fig:precip_timeseries} but for 5-day ahead forecasts.}
  \label{fig:precip_timeseries_lag5}
\end{figure}

\begin{figure}
  \centering
  \includegraphics[width=\linewidth]{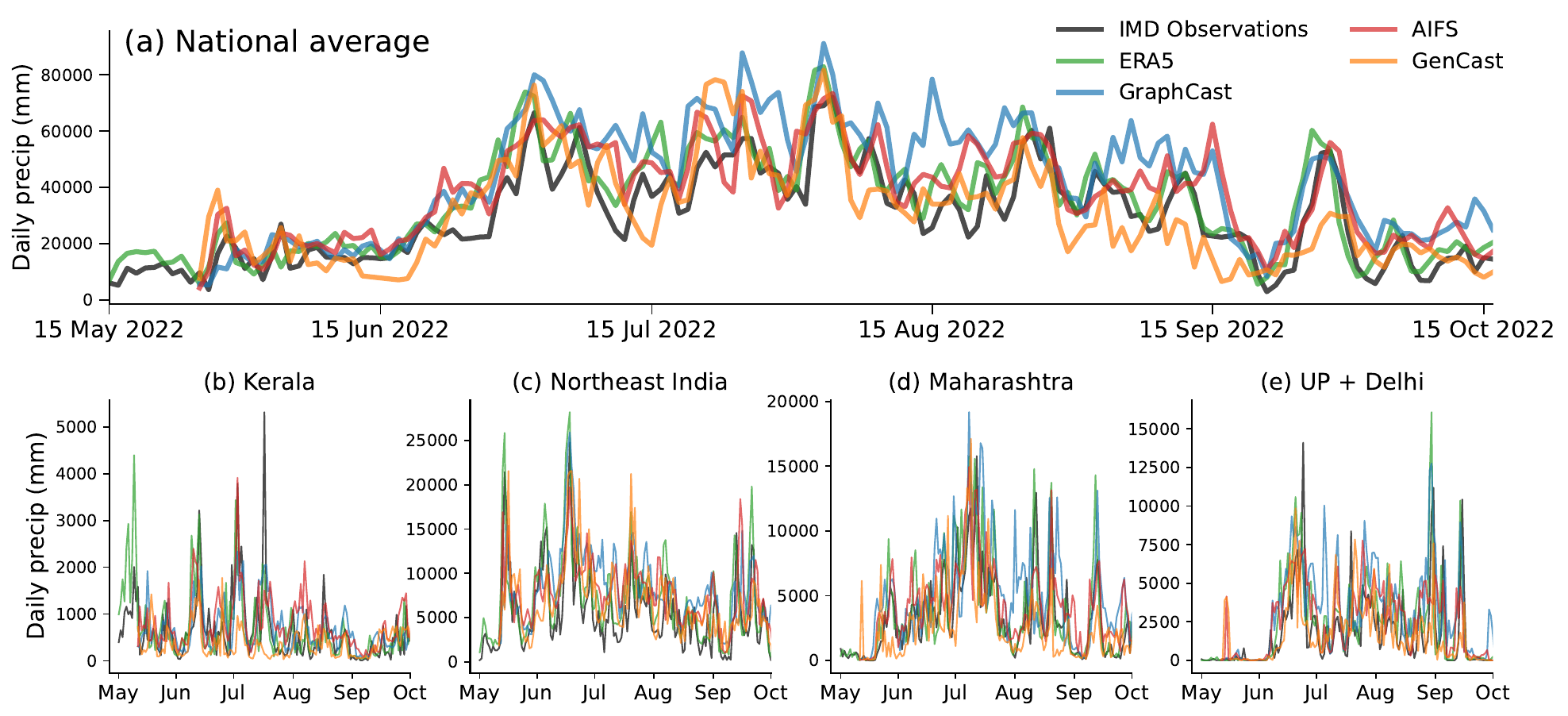}
  \caption{\small Same as Figure \ref{fig:precip_timeseries} but for 10-day ahead forecasts. At these lead times, the models severely underestimate extreme precipitation events. On the other hand, all models overpredict rainfall throughout the monsoon season, often even on dates when the actual precipitation was negligible.}
  \label{fig:precip_timeseries_lag10}
\end{figure}

\clearpage
\section{Ensemble Dispersion: GenCast vs. IFS Ensemble}

\begin{figure} [!ht]
  \centering
  \includegraphics[width=0.5\linewidth]{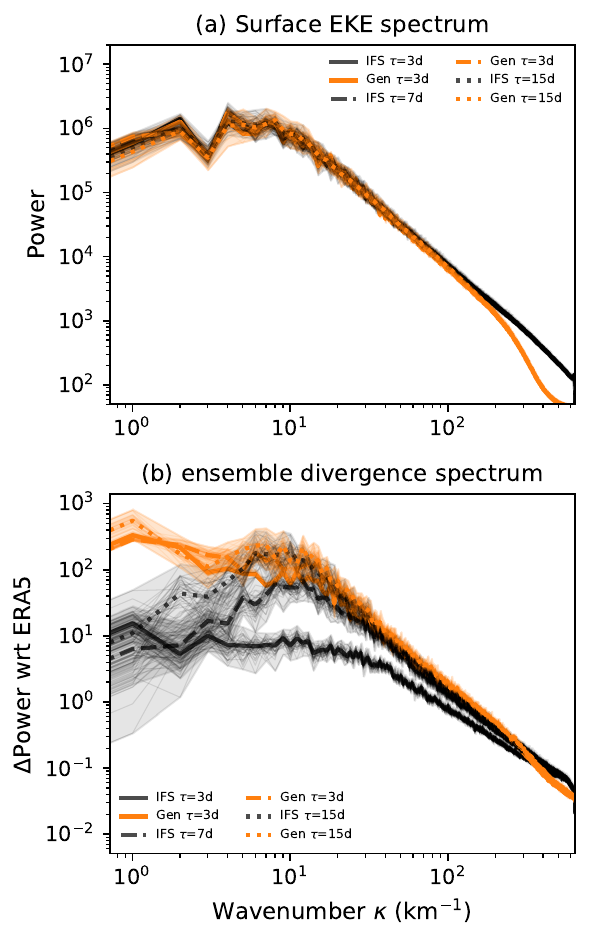}
  \caption{\small \textbf{Spectral analysis reveals distinct ensemble generation strategies between AI and traditional approaches.} (a) Surface eddy kinetic energy spectra and (b) ensemble divergence spectra comparing ECMWF IFS ensemble (black/gray lines) and GenCast ensemble (orange lines) forecasts during Cyclone Tauktae (May 2021) across forecast lead times $\tau = 3$, 7, and 15 days. Individual ensemble members are shown as thin lines with ensemble means as thick lines. GenCast demonstrates reasonable short-term uncertainty representation but exhibits distinct spectral characteristics at extended lead times, reflecting fundamental differences between physics-based and diffusion-based ensemble generation methodologies. The scale-dependent uncertainty patterns provide insights into how different ensemble approaches represent atmospheric predictability limits and chaotic error growth processes.}
  \label{fig:ensemble_spectrum}
\end{figure}

\clearpage
\section{Subseasonal-to-seasonal Forecast Statistics}

\begin{figure}[!ht]
    \centering
    \includegraphics[width=\linewidth]{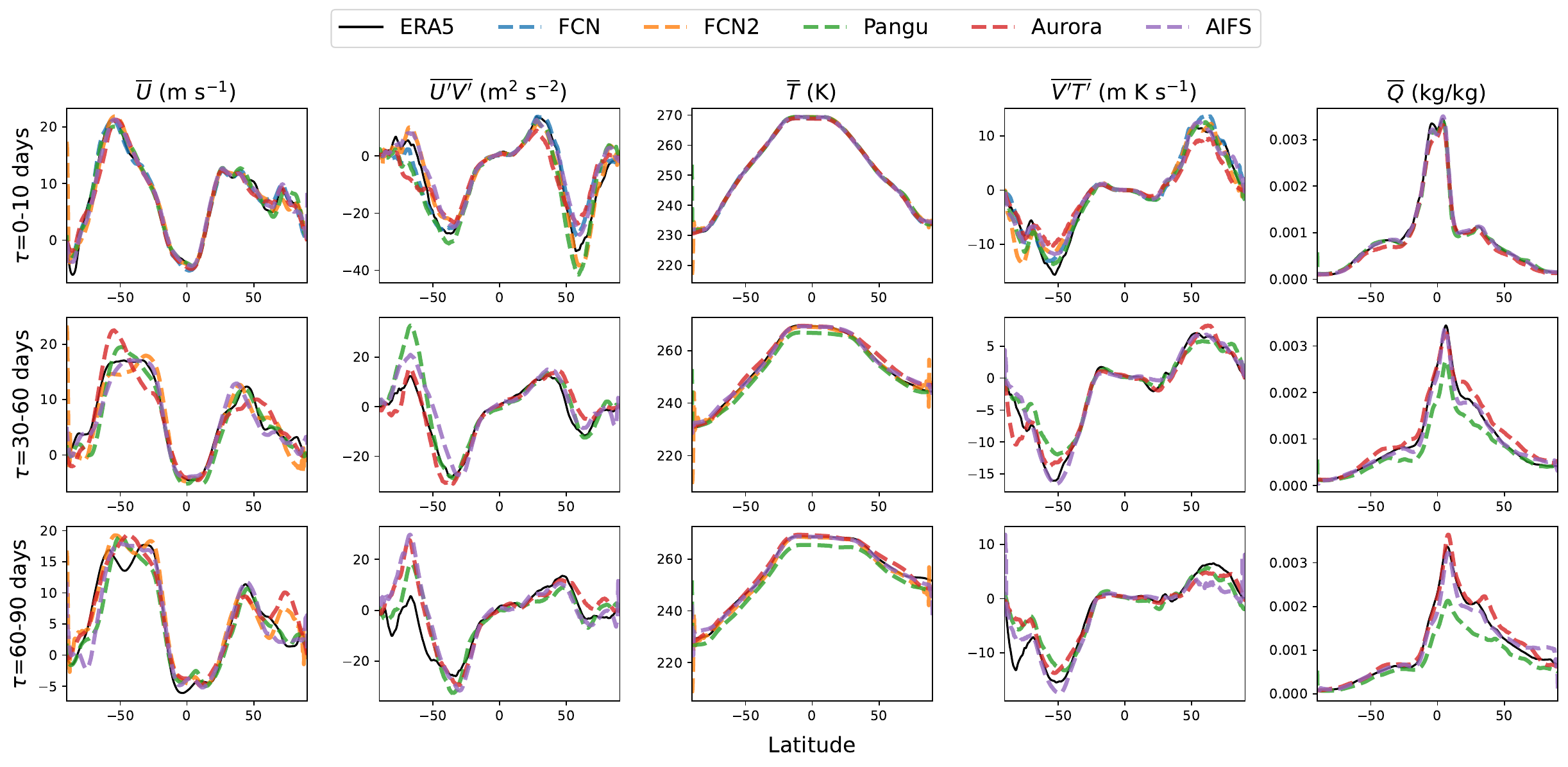}
    \caption{Zonal mean and eddy statistics for subseasonal-to-seasonal timescales, similar to Figure \ref{fig:s2s_structures}, but at 500 hPa. The models most notably diverge from the ERA5 statistics (in black) in the southern hemisphere. All models maintain a similar zonal mean temperature structure over all timescales, but exhibit markedly different zonal mean zonal winds, eddy momentum flux, eddy temperature flux, and the zonal mean specific humidity profile.}
    \label{fig:s2s_structure500hpa}
\end{figure}

\end{document}